\documentclass[aps,prl,reprint]{revtex4-1}

\usepackage{bm}
\usepackage{caption}
\usepackage{fullpage}
\usepackage{amsmath}
\usepackage{amssymb}
\usepackage{amsbsy}
\usepackage{graphicx}
\usepackage{indentfirst}
\usepackage{setspace}
\usepackage{color}
\usepackage{natbib}
\usepackage{cleveref}
\usepackage{subcaption}
\usepackage{epstopdf}
\graphicspath {{cascade/}}
\graphicspath {{visual/}}
\graphicspath {{spectrum/}}
\graphicspath {{vortex/}}
\graphicspath {{wave/}}
\newcommand{\be}{\begin{equation}}
\newcommand{\ee}{\end{equation}} 
\newcommand{\lb}{\label}
\newcommand{\OL}{\overline}

\newcommand{\inj}{{\scriptsize\mbox{inj}}}
\newcommand{\smax}{{\scriptsize\max}}

\newcommand{\bff}{{\bf f}}
\newcommand{\bk}{{\bf k}}

\newcommand{\br}{{\bf r}}

\newcommand{\bu}{{\bf u}}

\newcommand{\bx}{{\bf x}}

\newcommand{\bB}{{\bf B}}

\newcommand{\bJ}{{\bf J}}
\newcommand{\bS}{{\bf S}}

\newcommand{\mC}{{\mathcal C}^{ub}}
\newcommand{\mR}{{\mathcal R}^{ub}}

\newcommand{\bepsilon}{\pmb{\varepsilon}}
\newcommand{\btau}{\pmb{\tau}}

\newcommand{\grad}{{\mbox{\boldmath $\nabla$}}}
\newcommand{\bdot}{{\mbox{\boldmath $\cdot$}}}

\newcommand{\btimes}{{\mbox{\boldmath $\times$}}}

\begin{document}
\title{Decoupling of the Kinetic and Magnetic Energy Cascades in MHD Turbulence}
\author{Xin Bian}
\email{xin.bian@rochester.edu}
\affiliation{Department of Mechanical Engineering, University of Rochester}
\author{Hussein Aluie}
\affiliation{Department of Mechanical Engineering, University of Rochester}
\affiliation{Laboratory for Laser Energetics, University of Rochester}
\begin{abstract}
Magnetic and kinetic energy in ideal incompressible MHD are not global invariants and, therefore, it had been justified to discuss only the cascade of their sum, total energy. We provide a physical argument based on scale-locality of the cascade, along with compelling evidence that at high Reynolds numbers, magnetic and kinetic energy budgets statistically decouple beyond a transitional ``conversion'' range. This arises because magnetic field-line stretching is a large-scale process which vanishes on average at intermediate and small scales within the inertial-inductive range, thereby allowing each of mean kinetic and magnetic energy to cascade conservatively and at an equal rate. One consequence is that the turbulent magnetic Prandtl number is unity over the ``decoupled range'' of scales.
\end{abstract}

\maketitle

Magnetohydrodynamic (MHD) turbulence is of fundamental importance to many fields of science, including astrophysics, solar physics, space weather, and nuclear fusion. The Reynolds numbers of such flows are typically very large, giving rise to plasma fluctuations with power-law spectra over a vast range of scales where both viscosity and resistivity are negligible. We call such a range ``inertial-inductive'' since ideal dynamics dominate. There are several competing theories for the spectrum of strong MHD turbulence over the inertial-inductive range \cite{Iroshnikov64,Kraichnan65,GoldreichSridhar95,Boldyrev05,Boldyrev06}, all of which assume scale-locality of the energy cascade, which has been shown to hold \cite{AluieEyink10}.

In a scale-local cascade, energy transfer across scale $\ell$ is predominantly due to 
scales within a moderate multiple of $\ell$ \cite{Eyink05}.
This gives rise to an inertial-inductive scale range over which the flow evolves without 
direct communication with the largest or smallest scales in the system.

In MHD turbulence, only the sum of magnetic and kinetic energy (KE and ME, respectively), {\it i.e.} total energy, is a global invariant of the inviscid unforced dynamics. Therefore, it has been justified to discuss only the cascade of total energy, but not of KE or ME separately, which are coupled by magnetic field-line stretching. In principle, the process of magnetic field-line stretching can operate at \emph{all} scales, giving rise to various phenomena such as Alfv\'en waves.

We shall show here that  magnetic field-line stretching is a large-scale process, which operates over a ``conversion range'' of scales of limited extent and vanishes \emph{on average} at intermediate and small scales in the inertial-inductive range \cite{AluieThesis,*Aluie10}. Over the ensuing part of the inertial-inductive range, mean KE and ME cascade conservatively and at an equal rate to smaller scales despite not being separate invariants. 

Our findings are important in subgrid scale modeling of systems such as accretion disks, whose evolution is controlled by 
magnetic flux through the disk \cite{ShakuraSunyaev1973,BlandfordPayne1982,Hawleyetal95}. The strength of 
the magnetic field is determined by a balance between (i) turbulent advection (or turbulent viscosity) which accretes the field radially inward, and (ii) turbulent resistivity which diffuses it outward \cite{GuanGammie09,LesurLongaretti09,FromangStone09,Cao11}.
Other applications are outlined in the conclusion.

We start from the incompressible MHD equations with a constant density $\rho$:
\begin{align}
\partial _t \textbf u+ ( \textbf u\bdot \grad) \textbf u &=- \grad p+ \bJ \btimes \bB +\nu \nabla^2 \bu+ \bff,\label{velmhd}\\
\partial _t \textbf B &=\nabla\times(\textbf u\times \textbf B)+\eta\nabla^2 \textbf B. \label{magmhd}
\end{align}
Here $\bu$ is the velocity, and $\bB$ is the magnetic field normalized by $\sqrt{4\pi\rho}$ to have Alfv\'en (velocity) units. Both fields are solenoidal: $\grad\bdot\bu=\grad\bdot\bB=0$. The pressure is $p$, $\bJ=\grad\btimes\bB$ is (normalized) current density, $\textbf f$ is external forcing, $\nu$ is viscosity, and $\eta$ is resistivity.

In a statistically steady state, the space-averaged KE and ME budgets are, respectively,
\begin{align}
\langle S_{ij} B_i B_j \rangle=\epsilon^{\inj}-\nu\langle |\nabla \textbf u|^2\rangle, \label{energybud1}\\
\langle S_{ij} B_i B_j  \rangle=\eta\langle |\nabla \textbf B|^2\rangle, \label{energybud2}
\end{align}
where $\langle ... \rangle$ is a spatial average, $S_{ij}=(\partial_ju_i+\partial_iu_j)/2$ is the strain rate tensor, and $\epsilon^{\inj} = \langle\bff\bdot\bu\rangle$ is kinetic energy injection rate. It is clear from eqs. \eqref{energybud1}-\eqref{energybud2} that mean KE-to-ME conversion due to
magnetic field-line stretching is positive and bounded: $0\leq \langle \bB_{}\bdot\bS_{}\bdot\bB_{} \rangle\leq \epsilon^{\inj}$. The bound holds in the presence of an arbitrarily strong uniform magnetic field $\bB_0$, indicating significant cancellations. This can be understood by considering that in a turbulent flow, the strain $\bS$, being a derivative, is dominated by the small-scales, whereas $\bB$ is dominated by the large-scales, near the magnetic spectrum's peak, leading to decorrelation effects.

To analyze how magnetic field-line stretching operates at different length-scales, we utilize a coarse-graining approach for diagnosing multi-scale dynamics \cite{Eyink05,Aluie17}. 
A coarse-grained field which contains modes at length-scales $>\ell$ is defined by $\OL f_\ell(\bx) = \int d\br\, G_\ell(\br-\bx) f(\br)$, 
where $G_\ell(\br)\equiv \ell^{-3} G(\br/\ell)$ is a normalized kernel with its main weight in a ball of diameter $\ell$.
Coarse-grained MHD equations can then be written to describe $\OL\bu_\ell$ and $\OL\bB_\ell$, along with corresponding budgets 
for the quadratic invariants at scales $\ge\ell$, for \emph{arbitrary} $\ell$ in contrast to the mean field approach \cite{Moffatt78,KrauseRaedler80} (see \cite{Aluie17} and references therein). Hereafter, we drop subscript $\ell$ when possible.

KE and ME density balance at scales $>\ell$ are,
\begin{align}
\partial_t&(\frac{|\overline{\textbf {u}}|^2}{2})+\grad\bdot[\cdots]\nonumber\\
&=-\OL\Pi^u_{\ell}-\overline S_{ij} \overline B_i \overline B_j-\nu|\nabla \overline {\textbf u}|^2+\OL\bff\bdot\OL\bu,\label{kineticenergy} \\
\partial_t&(|\OL{\bB}|^2)+\grad\bdot[\cdots]\nonumber\\
&=-\OL\Pi^b_{\ell}+ \overline S_{ij} \overline B_i \overline B_j-\eta |\nabla \overline {\textbf B}|^2, \label{mageticenergy}
\end{align}
where $\grad\bdot[\cdots]$ represents spatial transport terms. 
Dissipation terms, $\nu|\nabla \overline {\textbf u}|^2$ and 
$\eta |\nabla \overline {\textbf B}|^2$, are mathematically guaranteed to be negligible \cite{Aluie17,Eyink18,*ZhaoAluie18} at scales $\ell\gg{\left(\ell_\nu,\ell_\eta\right)}$, with $\ell_\nu$ and $\ell_\eta$  the viscous and resistive length scales, respectively.

The first term on the RHS of eq.\eqref{kineticenergy}, $\OL\Pi^u_{\ell}$, appears as a sink in the KE budget of large scales $>\ell$ and
as a source in the KE budget of small scales $<\ell$ \cite{Aluie17}. It quantifies the KE transfer \emph{across} scale $\ell$, and is defined as 
$\OL\Pi^u_{\ell}\equiv-\OL{S}_{ij}\OL\tau_{ij}$, where $\OL\tau_{ij} = \tau_{\ell}(u_i, u_j)+\tau_{\ell}(B_i, B_j)$ is the sum of both the Reynolds and the Maxwell stress generated by scales $<\ell$ acting against the large-scale strain, $\OL{S}_{ij}$. Subscale stress is defined as $\tau_{\ell}(f, g)=\OL{\left(f g\right)}_{\ell}-\overline f_{\ell} \overline g_{\ell}$ for any two fields $f$ and $g$.
Similarly, $\OL\Pi^b_{\ell}\equiv-\OL{\bJ}_\ell\bdot \OL\bepsilon_\ell$ in eq. \eqref{mageticenergy} quantifies the ME transfer \emph{across} scale $\ell$, where $\OL\bepsilon_\ell \equiv \OL{\bu\btimes\bB} - \OL{\bu}\btimes\OL{\bB}$ is  (minus) the electric field generated by scales $<\ell$ acting on the large-scale current, $\OL\bJ=\grad\btimes\OL\bB$, resulting in a ``turbulent Ohmic dissipation'' to the small scales. 

Term $\OL\bB_{\ell}\bdot\OL\bS_{\ell}\bdot\OL\bB_{\ell}$ appears as a sink in eq.\eqref{kineticenergy} and a source in eq.\eqref{mageticenergy}, 
representing KE expended by the large-scale flow to bend and stretch large-scale $\OL\bB$-lines.
Unlike the cascade terms $\OL\Pi^u_\ell$ and $\OL\Pi^b_\ell$, which involve large-scale fields acting against subscale terms ($\OL\btau_\ell$ and $\OL\bepsilon_\ell$), $\OL\bB_{\ell}\bdot\OL\bS_{\ell}\bdot\OL\bB_{\ell}$ is purely due to large-scale fields and does not participate in energy transfer \emph{across} scale $\ell$. A more refined scale-by-scale analysis in \cite{AluieEyink10} showed how energy lost or gained from one field ($\bu$ or $\bB$) by line stretching reappeared in or disappeared from the other field at the \emph{same} scale.

In a steady state, space-averaging eqs. \eqref{kineticenergy},\eqref{mageticenergy} 
at any scale $\ell$ in the inertial-inductive range, $L\gg\ell\gg\left(\ell_\nu,\ell_\eta\right)$, yields
\begin{eqnarray}
\langle\overline \Pi^u_\ell\rangle&=&\epsilon^{\inj}-\mC(\ell), \label{eq:largeUbudget_steady}\\
\langle\overline \Pi^b_\ell\rangle&=&\mC(\ell),  \label{eq:largeMbudget_steady}
\end{eqnarray}
where we have dropped the dissipation terms and 
assumed that forcing is due to modes at scales $\sim L\gg\ell$, such that $\OL{\bff}_\ell = \bff$. Mean conversion, $\mC(\ell)\equiv\langle \overline S_{ij} \overline B_i \overline B_j \rangle$, in eqs. \eqref{eq:largeUbudget_steady},\eqref{eq:largeMbudget_steady} quantifies the cumulative KE-to-ME conversion at \emph{all} scales $>\ell$. 

Using scale-locality of the cascade terms, $\OL \Pi^u_\ell$ and $\OL \Pi^b_\ell$, which was proved in \cite{AluieEyink10}, we will now argue that mean magnetic field-line stretching is primarily a large-scale process which vanishes at intermediate and small scales within the inertial-inductive range. Note that the scale-locality discussed in \cite{Eyink05,AluieEyink10} is ``diffuse'' \cite{Kraichnan66} and states that contributions from disparate scales decay only as a power-law of the scale ratio.

Define $\ell_d$ as the largest scale at which non-ideal microphysics becomes significant, $\ell_d = \max(\ell_\nu,\ell_\eta)$.
Define the cumulative KE-to-ME conversion at scales $>\ell_d$ by $ \mC_d\equiv\mC(\ell_d)$, which is not necessarily equal to the unfiltered $\langle \bB_{}\bdot\bS_{}\bdot\bB_{} \rangle$ due to possible contributions from scales $<\ell_d$ [see discussion shortly after eq. \eqref{eq:MEbudget_Saturate} below].

Define $\ell_s$ as the largest scale at which $\mC(\ell_s)=\mC_d$. We'll argue that (i) $\ell_s \neq \ell_d$ and (ii) $\mC(\ell)=\mC_d$ for all
scales $\ell_s>\ell\gg\ell_d$.

First, assume $\ell_s$ = $\ell_d$. This implies that as functions of $\ell$, $\mC(\ell)=\langle \overline \Pi^b_\ell\rangle=\epsilon^{\inj}-\langle\overline \Pi^u_\ell\rangle$ depends on dissipative parameters $\nu$ or $\eta$. However, $\langle\overline \Pi^u_\ell\rangle$ and $\langle \overline \Pi^b_\ell\rangle$ are scale-local in the inertial-inductive range \cite{AluieEyink10} and are insensitive to the microphysics when $\ell\gg\ell_d$. Therefore, $\ell_s \neq \ell_d$. 
Second, if $\mC(\ell)\neq \mC_d$ over $\ell_s>\ell\gg\ell_d$, then $\mC(\ell)$, which we assume is continuous, will have an extremum at a scale $\ell_*$ within that range [since $\mC(\ell_s)=\mC(\ell_d)=\mC_d]$. Therefore, $\langle\overline \Pi^u_\ell\rangle$ and $\langle\overline \Pi^b_\ell\rangle$ will also have extrema, indicating the existence a special scale $\ell_*$ in the inertial-inductive range, in conflict with scale-invariance of the ideal MHD dynamics.

Therefore, $\mC(\ell) \rightarrow \mC_d$ within a conversion range $L>\ell>\ell_s$ and, over the ensuing range 
$\ell_s>\ell\gg\ell_d$, it saturates at $\mC(\ell)= \mC_d$. Since $\mC(\ell)$ measures the cumulative KE-to-ME conversion at all scales $>\ell$, saturation implies a zero contribution from $\ell_s>\ell\gg\ell_d$.
We conclude that mean KE-to-ME conversion, $\langle  S_{ij}  B_i  B_j \rangle$, is a large-scale process within the inertial-inductive range, acting over a conversion range $L>\ell>\ell_s$ of limited extent,  {\it i.e.} the scale-range does not increase asymptotically with the Reynolds number. Mean KE and ME budgets decouple in the absence of conversion over the ``decoupled range'' of scales, $\ell_s>\ell\gg\ell_d$:
\begin{eqnarray}
\langle\overline \Pi^u_\ell\rangle&=&\epsilon^{\inj}-\mC_d,\label{eq:KEbudget_Saturate}\\
\langle\overline \Pi^b_\ell\rangle&=&\mC_d.\label{eq:MEbudget_Saturate}
\end{eqnarray}
With the RHS of eqs.\eqref{eq:KEbudget_Saturate},\eqref{eq:MEbudget_Saturate} being independent of scale $\ell$, 
KE and ME each cascades conservatively after the mechanism coupling them halts. 
Scale-locality suggests that the normalized KE and ME cascade 
rates, $\langle\OL\Pi^u\rangle/\epsilon^\inj$ and $\langle\OL\Pi^b\rangle/\epsilon^\inj$, should have a universal value of order unity over 
$\ell_s>\ell\gg\ell_d$, regardless of the forcing, $Pr_m=\nu/\eta$, or $\bB_0$. Note that scale $\ell_s$ at which 
the budgets decouple is within the inertial-inductive range, despite the well-known non-equipartition of KE and ME spectra in that range \cite{MatthaeusGoldstein82,MatthaeusLamkin86,Grappinetal83,Boldyrevetal11} (Fig. 8 in SM).

While the above argument suggests that $\mC(\ell)$ should become constant at scales smaller than the conversion range, it only applies within the inertial-inductive range, $L\gg\ell\gg\ell_d$. It is possible for $\mC(\ell)$ to vary again when transitioning to scales $\lesssim\ell_d$. An example is the viscous-inductive (Batchelor) range, $\ell_\nu\gg\ell\gg\ell_\eta$, over which a scale-by-scale analysis in \cite{AluieEyink10} showed that magnetic field-line stretching can act as a forcing term in the ME budget, consistent with our understanding of high $Pr_m$ flows \cite{Zel'Dovichetal84,Schekochihinetal02}. The above argument for saturation of  $\mC(\ell)$ breaks down at scales $\lesssim\ell_d$, such as in the viscous-inductive range where scale-locality does not hold due to a smooth velocity field \cite{AluieEyink10}.

Our conclusions are supported by a suite of pseudospectral Direct Numerical Simulations (DNS) up to $2{,}048^3$ in resolution with phase-shift dealiasing, 
using hyperdiffusion and other parameters summarized in Table \ref{Tbl:Simulations}.

\begin{table}[]
\caption{Each suite of Runs was carried out at different Reynolds numbers at $256^3$, $512^3$, and $1{,}024^3$ resolutions. Run V was also conducted at $2{,}048^3$ resolution. $Pr_m=\nu/\eta$ is magnetic Prandtl number. $B^\smax_k=\sqrt{\max_k[E^b(k)]}$ is at the magnetic spectrum's [$E^b(k)$] peak. ABC (helical) and TG (non-helical) forcing were applied at wavenumber $k_f$. More details are in the supplemental material (SM).}
\begin{center}
\begin{tabular}{lcccccc}
\hline
\hline
Run   & Forcing        & $k_f$ & $Pr_m$ & $|\bB_0|/B^\smax_k$  \\ \hline
   $\rm I$     & ABC           & 2    & 1      & 0     \\
   $\rm II$     & ABC          & 2    & 1      & 10   \\
   $\rm III$    & TG & 1    & 1     & 0    \\
   $\rm IV$    & ABC          & 1    & 2     & 0    \\ 
   $\rm V$    & ABC          & 2    & 1     & 2    \\ 
\hline
\hline
\end{tabular}
\end{center}
\lb{Tbl:Simulations}\end{table}

\begin{figure*}
\centering
\vspace{-0.5cm}
\begin{subfigure}{0.32\textwidth}
\includegraphics[width=2.2 in]{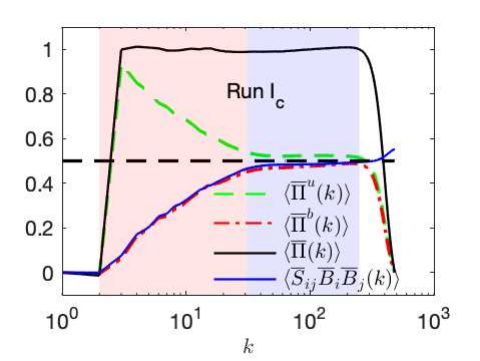}
\end{subfigure}
\begin{subfigure}{0.32\textwidth}
\includegraphics[width=2.2 in]{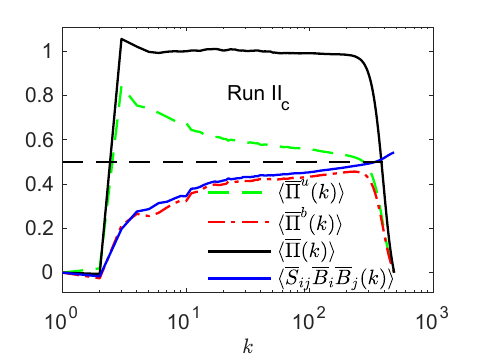}
\end{subfigure}
\begin{subfigure}{0.32\textwidth}
\includegraphics[width=2.2 in]{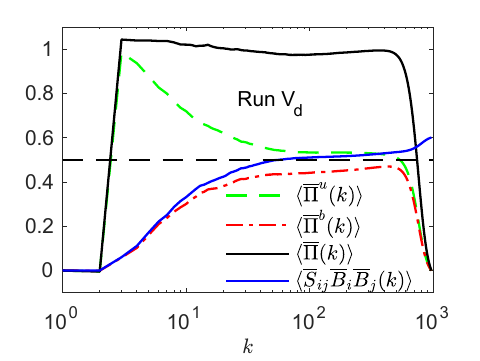}
\end{subfigure}
\\
\vspace{-0.1cm}
\begin{subfigure}{0.32\textwidth}
\includegraphics[width=2.2 in]{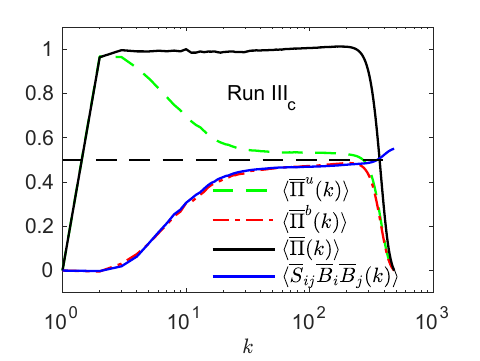}
\end{subfigure}
\begin{subfigure}{0.32\textwidth}
\includegraphics[width=2.2 in]{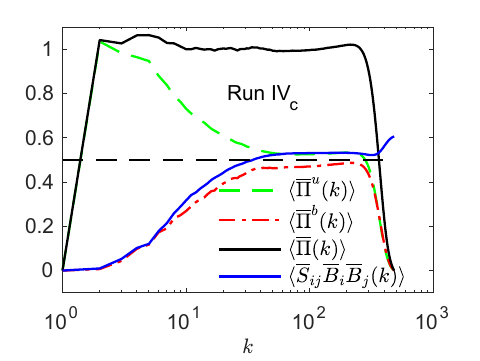}
\end{subfigure}
\begin{subfigure}{0.32\textwidth}
\includegraphics[width=2.2 in]{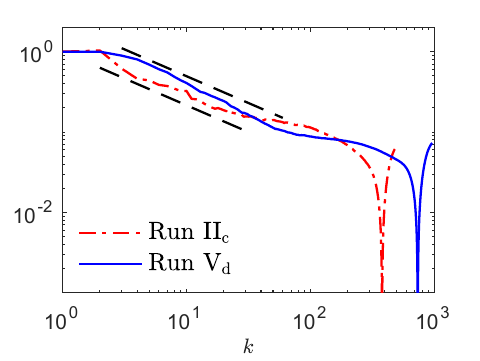}
\end{subfigure}
\renewcommand{\figurename}{FIG.}
\vspace{-0.2cm}
\caption{The first five panels show $\langle \overline \Pi\rangle=\langle\overline \Pi^u+\overline \Pi^b\rangle$, $\langle \overline \Pi^u\rangle$, $\langle\overline \Pi^b\rangle$, and $\langle\overline {S}_{ij}\overline{B}_i\overline{B}_j\rangle$ as a function of $k\equiv2\pi/\ell$ from our  highest resolution Runs ($1{,}024^3$ for Runs I to IV and $2{,}048^3$ for Run V. See SM for lower resolutions).  In top-left panel, conversion  (decoupled) range is shaded red (blue). All plots are time-averaged and normalized by $\epsilon^{\inj}$. The horizontal straight dashed line is at 0.5. Bottom-right panel shows a log-log plot of relative residual conversion, $\mR(k)/\mC_d$, and a reference black-dashed line with a $-2/3$ slope, suggesting that KE-to-ME conversion saturates in a manner consistent with scale-locality \cite{AluieEyink10}.
}
\label{fig:mhdflux}
\end{figure*}

Figure \ref{fig:mhdflux} shows results from the five flows we consider, at the highest resolution (see SM for lower resolution runs and evidence of convergence). In all runs, total energy, 
being a global invariant, is transferred conservatively across scales $L\gg\ell\gg \ell_d$, as indicated by a scale-independent total energy flux,
$\langle \overline \Pi_\ell\rangle=\langle\overline \Pi^u_\ell+\overline \Pi^b_\ell\rangle$. Both $\OL\Pi^u_\ell$ and $\OL\Pi^b_\ell$ decay to zero at scales $\lesssim\ell_d$, when the nonlinearities shut down in the dissipation range. Mean KE-to-ME conversion, $\mC(\ell)$, increases from 0
at the largest scales to $\approx \mC_d \approx \epsilon^\inj/2$ at an intermediate scale $\ell_s$ within the inertial-inductive range. 
Over the ensuing range, $\ell_s>\ell\gg\ell_d$, $\mC(\ell)$ is scale-independent, indicating a negligible contribution to magnetic field-line stretching at these scales. There is a slight increase in $\mC(\ell)$ in the dissipation range, at scales $\lesssim\ell_d$ where our argument is not expected to hold due to a lack of scale-locality.
In all cases, $\langle\OL \Pi^b_\ell\rangle \approx \mC(\ell)$ and $\OL \Pi^u_\ell \approx \epsilon^\inj - \mC(\ell)$ over the inertial-inductive range, consistent with eqs. \eqref{eq:largeUbudget_steady},\eqref{eq:largeMbudget_steady}.
Beyond the conversion range, scale-transfer becomes independent of $\ell$, $\langle\OL \Pi^u_\ell \rangle\approx \epsilon^\inj - \mC_d$ and $\langle\OL \Pi^b_\ell \rangle\approx  \mC_d$ over $\ell_s>\ell\gg\ell_d$, consistent with eqs. \eqref{eq:KEbudget_Saturate},\eqref{eq:MEbudget_Saturate}, and indicative of a conservative cascade of KE and ME energy, respectively.
In all runs, we observe  that the KE and ME cascade rates become equal in magnitude, $\langle\OL \Pi^u_\ell\rangle\approx\langle\OL \Pi^b_\ell\rangle$, over $\ell_s>\ell\gg\ell_d$, with magnetic field-line stretching channeling $\approx 1/2$ of the injected energy to the magnetic field,
regardless of the forcing, $Pr_m$, or $\bB_0$, consistent with scale-locality.

Among the five cases in Fig. \ref{fig:mhdflux}, the conversion range is widest in the presence of $|\bB_0|/B^\smax_k=10$ (Run $\rm II_c$). However, according to our argument, its extent cannot increase indefinitely with an increasing dynamic range of scales (or Reynolds number, $Re$).
After all, $\langle \bB\bdot\bS\bdot\bB\rangle$ is bounded even in the $|\bB_0|\to\infty$ limit.
Indeed, a plot of the relative residual conversion  $\mR(\ell)/\mC_d \equiv \langle \OL\bB_{\ell_d}\bdot\OL\bS_{\ell_d}\bdot\OL\bB_{\ell_d} -\OL\bB_{\ell}\bdot\OL\bS_{\ell}\bdot\OL\bB_{\ell} \rangle /\langle \OL\bB_{\ell_d}\bdot\OL\bS_{\ell_d}\bdot\OL\bB_{\ell_d} \rangle$ in Fig. \ref{fig:mhdflux} (and Fig. 5 in SM) decays at least as fast as a power-law as $\ell\to\ell_d$, consistent with what is expected from scale-locality (we take $\ell_d$ as the scale at which $\langle \overline \Pi_\ell\rangle = \epsilon^\inj/2$). Moreover,
plots of $\mC(\ell)$ at increasing $Re$ (Fig. 4 in SM) show a clear convergence to $\mC_d\approx\epsilon^\inj/2$.

\begin{figure*}
\centering
\begin{subfigure}{0.45\textwidth}
\includegraphics[width=2.2 in]{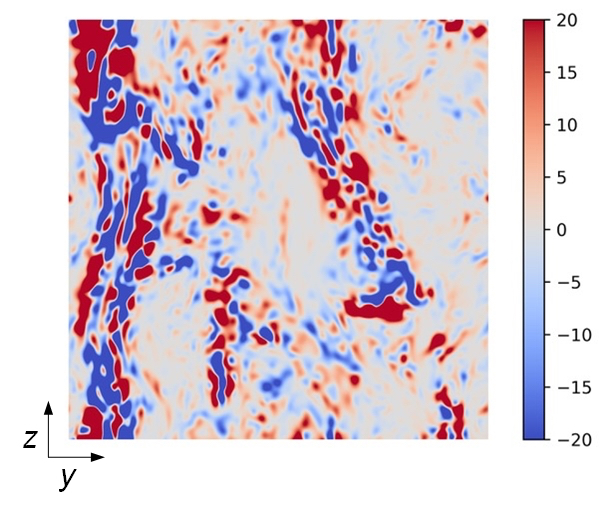}
\end{subfigure}
\begin{subfigure}{0.45\textwidth}
\includegraphics[width=2.2 in]{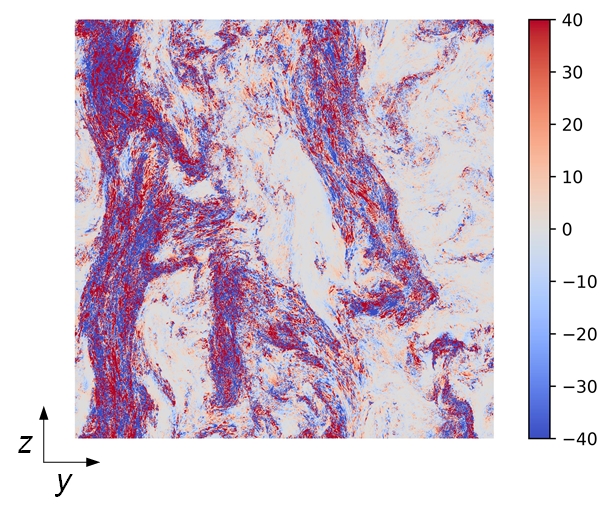}
\end{subfigure}
\\
\centering
\begin{subfigure}{0.45\textwidth}
\includegraphics[width=2.2 in]{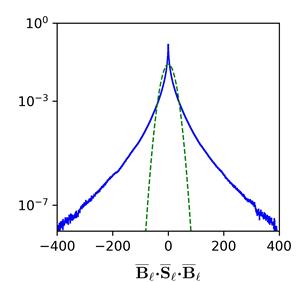}
\end{subfigure}
\begin{subfigure}{0.45\textwidth}
\includegraphics[width=2.2 in]{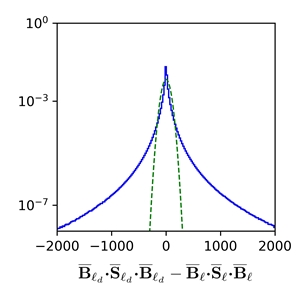}
\end{subfigure}
\renewcommand{\figurename}{FIG.}
\caption{For scale $\ell=2\pi/30$ ($k=30$) from Run $\rm II_c$ in Figure \ref{fig:mhdflux} at one instant in time: top two panels show a 2D slice from the 3D domain of pointwise conversion at large scales, $\OL\bB_{\ell}\bdot\OL\bS_{\ell}\bdot\OL\bB_{\ell} (\bx)$ (top left), and small scales, $\OL\bB_{\ell_d}\bdot\OL\bS_{\ell_d}\bdot\OL\bB_{\ell_d} (\bx) -\OL\bB_{\ell}\bdot\OL\bS_{\ell}\bdot\OL\bB_{\ell} (\bx)$  (top right). $\bB_0$ is in the z-direction. 
Bottom two panels show probability density function of conversion as a function of $\bx$ at large scales (bottom left) and small scales (bottom right). The large-scale distribution has mean of 0.43 and variance of 223.54. The small-scale distribution has mean of 0.09 and variance of 3060.84. Quantities are normalized by energy injection rate $\epsilon^{\inj}$. Unnormalized Gaussians (green dashed lines) are added to both plots. }
\label{fig:mhdvis}
\end{figure*}

The negligible mean KE-to-ME conversion at small scales within the decoupled range might seem counterintuitive at first. After all, a hallmark of MHD turbulence are Alfv\'en waves which are fastest at small scales. The decoupling of ME and KE budgets poses no contradiction since it is only in the \emph{mean}, which allows for decorrelation effects at small scales similar to those arising in compressible turbulence \cite{Aluie11,*Aluieetal12,Yangetal16}. Utilizing the simultaneous information in both scale and space
afforded by our coarse-graining approach, we analyze $\OL\bB_{\ell}\bdot\OL\bS_{\ell}\bdot\OL\bB_{\ell} (\bx)$ acting on scales $>\ell$ and the residual conversion within the inertial-inductive range, $\OL\bB_{\ell_d}\bdot\OL\bS_{\ell_d}\bdot\OL\bB_{\ell_d} (\bx) -\OL\bB_{\ell}\bdot\OL\bS_{\ell}\bdot\OL\bB_{\ell} (\bx)$, as a function of space $\bx$ in Fig. \ref{fig:mhdvis}. 
For an intermediate scale $\ell=2\pi/30$ from Run $\rm II_c$ (and Run $\rm I_c$ in Fig. 7 of SM), Fig. \ref{fig:mhdvis} shows how magnetic field-line stretching, which is concentrated in magnetic filaments, is an order of magnitude more intense at scales smaller than $\ell=2\pi/30$ compared to larger scales. Yet, the small-scale contribution fluctuates vigorously in sign, yielding a mere $17\%$ ($10\%$ in Run $\rm I_c$ in Fig. 7 of SM) to the space average. 
To illuminate the role of waves, we repeat in the SM the analysis above on two examples of non-colliding Alfv\'en waves, a monochromatic wave and a wavepacket, which are exact solutions of the MHD equations and which lack energy transfer between scales.

In conclusion, small-scales of the magnetic field in the decoupled scale range are maintained, on average, by turbulent Ohmic dissipation (the ME cascade), $\langle\OL\Pi^b\rangle=\langle\OL\bJ\bdot\OL\bepsilon\rangle$. Mean magnetic field-line stretching acts as a large-scale driver of the ME cascade, justifying the inclusion of a low-mode forcing in the induction eq. \eqref{magmhd} when resolving the transitional conversion range is unimportant, such as in high-$Re$ asymptotic scaling studies of MHD turbulence \cite{Masonetal08,Matthaeusetal08,Krstulovicetal14}. Our results will help in deriving relations equivalent to the Politano-Pouquet relations \cite{PolitanoPouquet98a,*PolitanoPouquet98b} but for the separate cascades of KE and ME, with potential implications on the scaling in MHD turbulence. This work can also help sub-grid scale model development and testing in Large Eddy Simulations of MHD turbulence \cite{MuellerCarati02,Mieschetal15}. For example, they provide a direct measure of the turbulent magnetic Prandtl number, which is unity within decoupled range due to equipartition of the cascades, $\langle \overline \Pi^u_\ell\rangle=\langle \overline \Pi^b_\ell\rangle$, which has important implications to astrophysical flows such as in accretion disks \cite{GuanGammie09,FromangStone09,Cao11}.
Our findings are also relevant for turbulent magnetic reconnection \cite{MatthaeusLamkin86,LazarianVishniac99,Eyinketal13} since they imply that the net bending and twisting of magnetic field lines at length scales in the decoupled range is driven by the effective electric field, $-\OL\bepsilon_\ell$, rather than by the flow's strain, giving independent support to previous studies \cite{MatthaeusLamkin86,Eyink15,*EyinkAluie06}.
Our framework for quantifying field-line stretching at various scales may also prove insightful in future studies of the magnetic dynamo \cite{BlackmanField02,Beresnyak12,Brandenburg18,Offermansetal18}.

\section*{Acknowledgements}
\noindent HA thanks G. Eyink for discussions. This work was supported by the DOE FES grant DE-SC0014318. HA was also partially supported by NASA grant 80NSSC18K0772, DOE grant DE-SC0019329, and NNSA award DE-NA0003856. Computing time was provided by NERSC under Contract No. DE-AC02-05CH11231.

\newpage
\renewcommand{\theequation}{A-\arabic{equation}}
\setcounter{equation}{0}  % reset counter 
\section*{Supplementary Material}  % use *-form to suppress numbering

The supplementary material provides simulation details are not included in the main Letter, and two examples to illustrate the role of waves.

\subsection{Numerical Setup}
Our numerical simulations of mechanically forced turbulence are conducted in a periodic box $\mathbb{T}^3=[0, 2\pi)^3$ with mesh resolution of $N^3$ grid points using a pseudospectral code. We use phase-shift dealiasing, which allows for $\approx 30\%$ increase in the dynamic range of scales compared to $2/3$-rd dealiasing \cite{PattersonOrszag71}. Time integration uses a second-order Adam-Bashforth scheme. 

We use hyperviscosity and hyperresistivity \cite{BorueOrszag95} commonly used in MHD turbulence studies  \cite{ChoVishniac00,Kawai13,Beresnyak15,Meyrandetal16,Kawazuraetal18} to reduce the dissipation range extent, thereby allowing for a longer inertial-inductive range of scales. Since our results pertain to 3rd-order moments in the form of energy transfer (which are cubic quantities) appearing in the energy budgets, they are not affected by hyperdiffusion which has been known to affect other (than 3rd-order) moments at high-wavenumbers such as an enhanced ``bottle-neck'' in the spectrum  \cite{BiskampMuller00,Frischetal08} and reduced intermittency \cite{Spyksmaetal12}. Unlike in non-conducting flows, normal viscosity and Spitzer resistivity in MHD turbulence are themselves rudimentary models of transport which do not faithfully capture the kinetic plasma physics (e.g. collisionless damping \cite{Lietal16}) and more self-consistent models such as Braginskii's treatment \cite{Braginskii65} are often needed in applications (see also \cite{Haines86,Daviesetal15}).
We use a Laplacian with an exponent $\alpha=5$, thus $\nu \nabla^2 \bu$ and $\eta\nabla^2 \bB$ in eqs. (1),(2) in the main Letter are replaced by $-\nu_h(-\nabla^2)^{\alpha} \bu $ and $-\eta_h(-\nabla^2)^{\alpha} \bB$, respectively, where $\nu_h$ is hyperviscosity, and $\eta_h$ is hyperresistivity coefficients. 

Runs I, II, IV, and V (see Table \ref{Tbl:Parameters} below) are driven by ABC forcing, which is helical:
\begin{equation}
\begin{aligned}
\textbf f & \equiv [A\sin(k_f z)+C\cos(k_f y) ]\textbf e_x +[B\sin(k_f x) \\
& + A\cos(k_f z) ]\textbf e_y + [C\sin(k_f y)+B\cos(k_f x) ]\textbf e_z,
\end{aligned}
\end{equation}
where $A=B=C=0.25$, $k_f$ is forcing wavenumber, $\textbf e_x$, $\textbf e_y$, and $\textbf e_z$ are unit vectors in $x$, $y$, and $z$, respectively. Taylor-Green (TG) forcing, which is non-helical, is used to drive the flow in Run III:
\begin{equation}
\begin{aligned}
\textbf f \equiv & f_0[\sin(k_f x)\cos(k_f y)\cos(k_f z) \textbf e_x\\
&-\cos(k_f x) \sin(k_f y)\cos(k_f z)\textbf e_y],
\end{aligned}
\end{equation}
where the force amplitude $f_0=0.25$. The simulation parameters are shown in Table \ref{Tbl:Parameters} below, where subscripts a, b, c, and d (e.g. Run $\rm V_a$ vs. Run $\rm V_b$ vs. Run $\rm V_c$ vs. Run $\rm V_d$) denote simulations using the same parameters but at different grid resolution (or Reynolds number).

\begin{table*}[]
\centering
\caption{Simulations parameters: $Pr_m$ is magnetic Prandtl number. $B^\smax_k=\sqrt{\max_k[E^b(k)]}$ is at the magnetic spectrum's [$E^b(k)$] peak. ABC (helical) and TG (non-helical) forcing were applied at wavenumber $k_f$.
}
%\label{simpara}
%\begin{tabular}{lllllllll}
\begin{tabular}{lcccccccc}
\hline
\hline
Run  & Grid  & Forcing                    & $k_f$ & $Pr_m$ & $|\bB_0|/B^\smax_k$        & $\nu_h$   & $\eta_h$ \\ \hline
   $\rm I_a$     & $256^3$  & ABC           & 2    & 1      & 0   & $5\times 10^{-16}$  & $5\times 10^{-16}$ \\
   $\rm I_b$     & $512^3$  & ABC           & 2    & 1      & 0   & $2\times 10^{-21}$  & $2\times 10^{-21}$  \\
   $\rm I_c$     & $1{,}024^3$  & ABC       & 2    & 1      & 0   & $4\times 10^{-25}$  & $4\times 10^{-25}$   \\
   $\rm II_a$    & $256^3$  & ABC           & 2    & 1      & 10  & $5\times 10^{-16}$  & $5\times 10^{-16}$ \\
   $\rm II_b$   & $512^3$  & ABC            & 2    & 1      & 10  & $2\times 10^{-21}$  & $2\times 10^{-21}$   \\
   $\rm II_c$   & $1{,}024^3$  & ABC        & 2    & 1      & 10  & $4\times 10^{-25}$  & $4\times 10^{-25}$ \\
   $\rm III_a$   & $256^3$  & TG            & 1    & 1      & 0   & $5\times 10^{-16}$  & $5\times 10^{-16}$  \\
   $\rm III_b$  & $512^3$  & TG             & 1    & 1      & 0   & $2\times 10^{-21}$  & $2\times 10^{-21}$    \\
   $\rm III_c$  & $1{,}024^3$  & TG         & 1    & 1      & 0   & $4\times 10^{-25}$  & $4\times 10^{-25}$    \\
   $\rm IV_a$    & $256^3$  & ABC           & 1    & 2      & 0   & $2\times 10^{-16}$  & $1\times 10^{-16}$ \\ 
   $\rm IV_b$    & $512^3$  & ABC           & 1    & 2      & 0   & $4\times 10^{-21}$  & $2\times 10^{-21}$   \\ 
   $\rm IV_c$    & $1{,}024^3$  & ABC       & 1    & 2      & 0   & $4\times 10^{-25}$  & $2\times 10^{-25}$  \\
   $\rm V_a$     & $256^3$  & ABC           & 2    & 1      & 2   & $5\times 10^{-16}$  & $5\times 10^{-16}$ \\
   $\rm V_b$     & $512^3$  & ABC           & 2    & 1      & 2   & $2\times 10^{-21}$  & $2\times 10^{-21}$  \\
   $\rm V_c$     & $1{,}024^3$  & ABC       & 2    & 1      & 2   & $4\times 10^{-25}$  & $4\times 10^{-25}$   \\ 
   $\rm V_d$    & $2{,}048^3$  & ABC        & 2    & 1      & 2   & $1\times 10^{-27}$  & $1\times 10^{-27}$  \\
\hline
\hline
\end{tabular}
\label{Tbl:Parameters}
\end{table*}

In the main Letter, we present results only at the highest resolution from Runs I-V. Here, we provide results at all resolutions from Runs I-V. 

\begin{figure*}
\centering
\begin{subfigure}{0.32\textwidth}
\includegraphics[width=2.2 in]{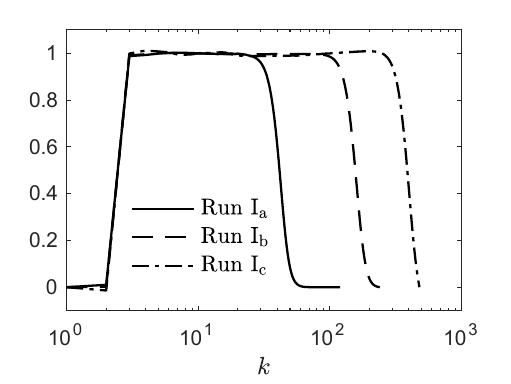}
\label{}
\end{subfigure}
\begin{subfigure}{0.32\textwidth}
\includegraphics[width=2.2 in]{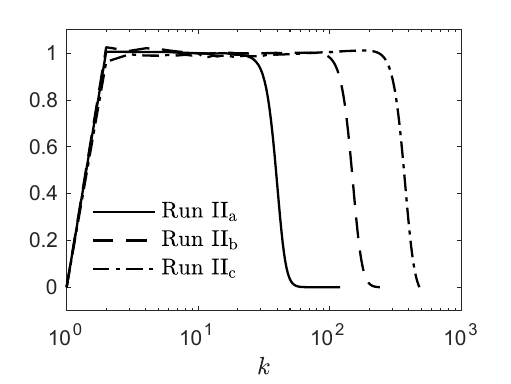}
\label{}
\end{subfigure}
\\
\vspace{-0.1cm}
\begin{subfigure}{0.32\textwidth}
\includegraphics[width=2.2 in]{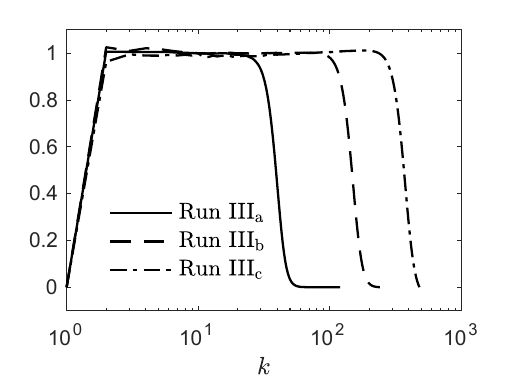}
\label{}
\end{subfigure}
\begin{subfigure}{0.32\textwidth}
\includegraphics[width=2.2 in]{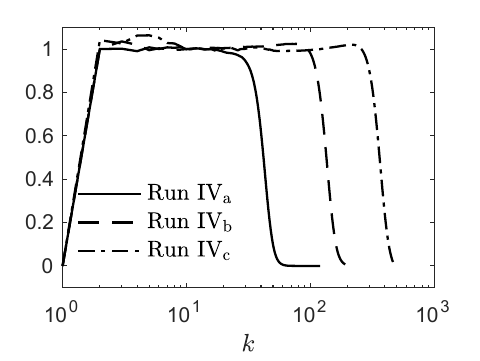}
\label{}
\end{subfigure}
\begin{subfigure}{0.32\textwidth}
\includegraphics[width=2.2 in]{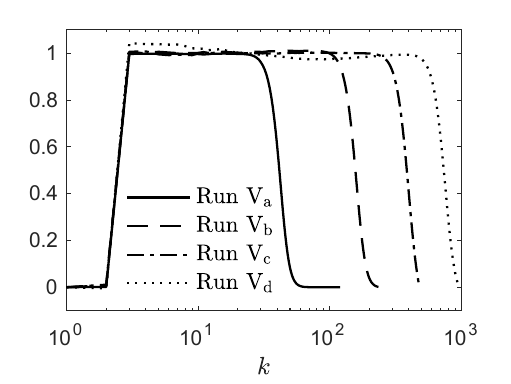}
\label{}
\end{subfigure}
\renewcommand{\figurename}{FIG.}
\caption{Plots showing convergence of $\langle \overline \Pi_\ell\rangle=\langle\overline \Pi^u_\ell+\overline \Pi^b_\ell\rangle$ by increasing resolution from $256^3$ (solid line) to $512^3$ (dashed line) to $1{,}024^3$ (dot dashed line) to $2{,}048^3$ (dotted line, only applicable to Run V).}
\label{totalfluxConverg}
\end{figure*}

\begin{figure*}
\centering
\begin{subfigure}{0.32\textwidth}
\includegraphics[width=2.2 in]{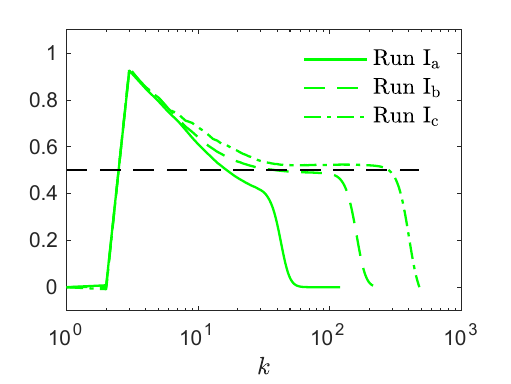}
\label{}
\end{subfigure}
\begin{subfigure}{0.32\textwidth}
\includegraphics[width=2.2 in]{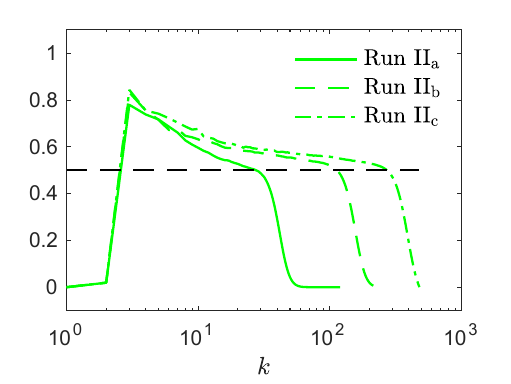}
\label{}
\end{subfigure}
\\
\vspace{-0.1cm}
\begin{subfigure}{0.32\textwidth}
\includegraphics[width=2.2 in]{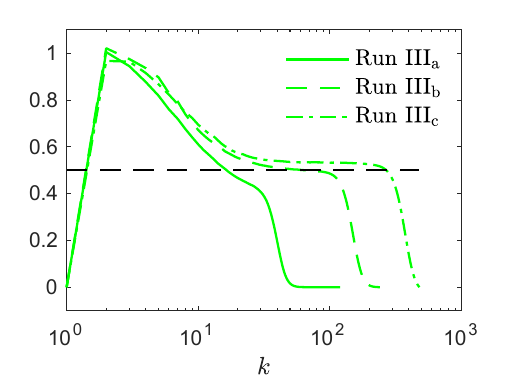}
\label{}
\end{subfigure}
\begin{subfigure}{0.32\textwidth}
\includegraphics[width=2.2 in]{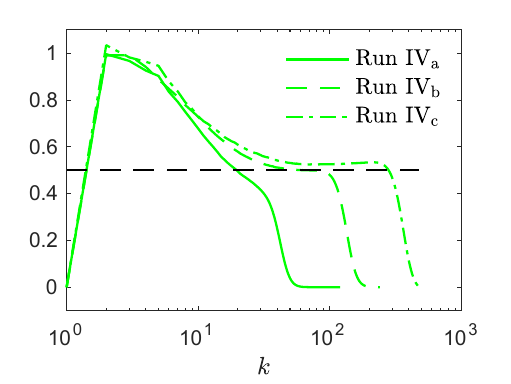}
\label{}
\end{subfigure}
\begin{subfigure}{0.32\textwidth}
\includegraphics[width=2.2 in]{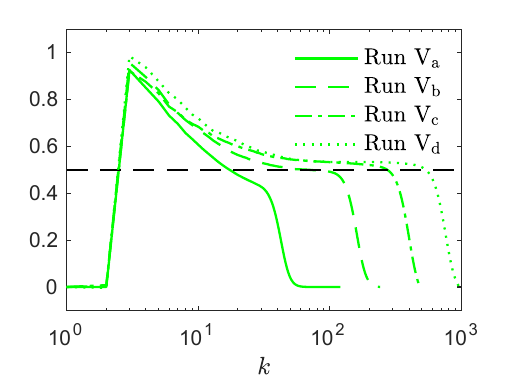}
\label{}
\end{subfigure}
\renewcommand{\figurename}{FIG.}
\caption{Plots showing convergence of $\langle\overline \Pi^u_\ell\rangle$ by increasing resolution from $256^3$ (solid line) to $512^3$ (dashed line) to $1{,}024^3$ (dot dashed line) to $2{,}048^3$ (dotted line, only applicable to Run V).}
\label{ufluxConverg}
\end{figure*}

\begin{figure*}
\centering
\begin{subfigure}{0.32\textwidth}
\includegraphics[width=2.2 in]{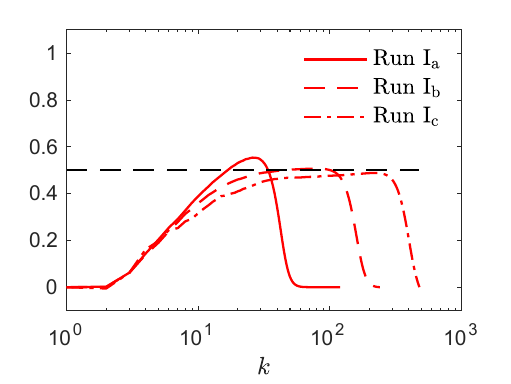}
\label{}
\end{subfigure}
\begin{subfigure}{0.32\textwidth}
\includegraphics[width=2.2 in]{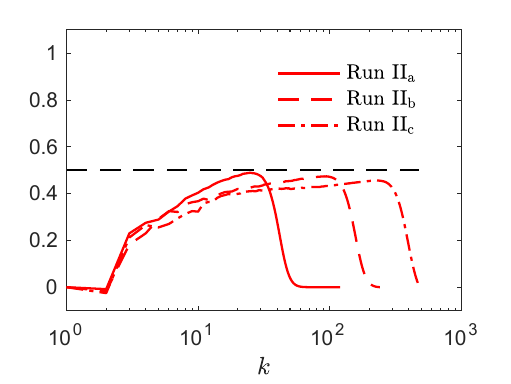}
\label{}
\end{subfigure}
\\
\vspace{-0.1cm}
\begin{subfigure}{0.32\textwidth}
\includegraphics[width=2.2 in]{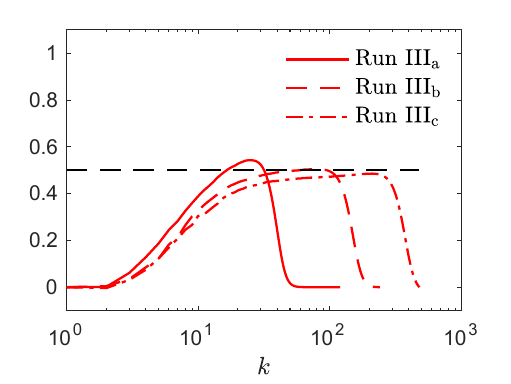}
\label{}
\end{subfigure}
\begin{subfigure}{0.32\textwidth}
\includegraphics[width=2.2 in]{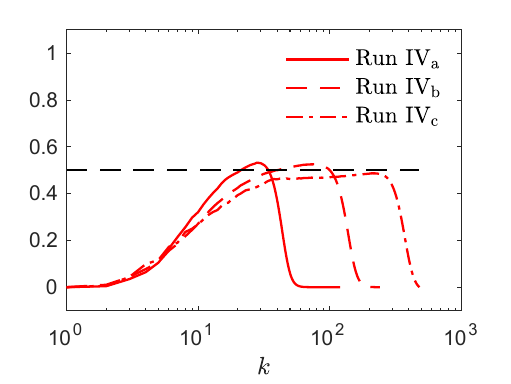}
\label{}
\end{subfigure}
\begin{subfigure}{0.32\textwidth}
\includegraphics[width=2.2 in]{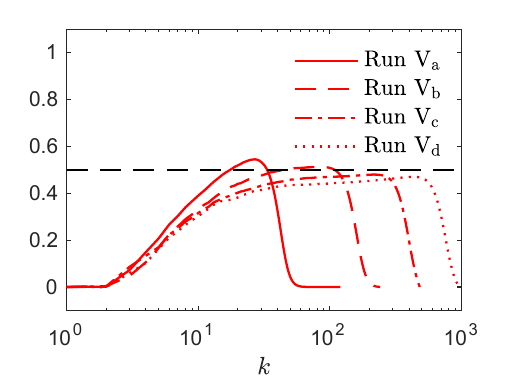}
\label{}
\end{subfigure}
\renewcommand{\figurename}{FIG.}
\caption{Plots showing convergence of $\langle\overline \Pi^b_\ell\rangle$ by increasing resolution from $256^3$ (solid line) to $512^3$ (dashed line) to $1{,}024^3$ (dot dashed line) to $2{,}048^3$ (dotted line, only applicable to Run V).}
\label{mfluxConverg}
\end{figure*}

\begin{figure*}
\centering
\begin{subfigure}{0.32\textwidth}
\includegraphics[width=2.2 in]{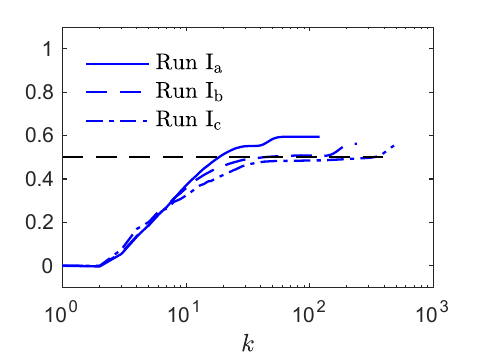}
\label{}
\end{subfigure}
\begin{subfigure}{0.32\textwidth}
\includegraphics[width=2.2 in]{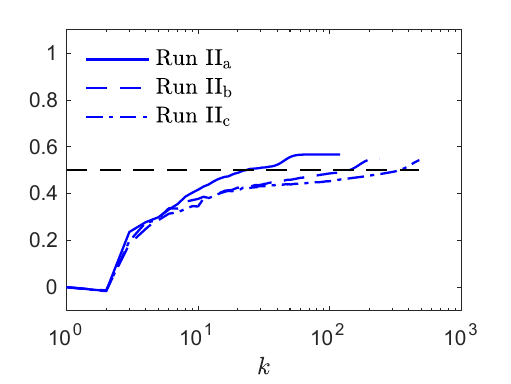}
\label{}
\end{subfigure}
\\
\vspace{-0.10cm}
\begin{subfigure}{0.32\textwidth}
\includegraphics[width=2.2 in]{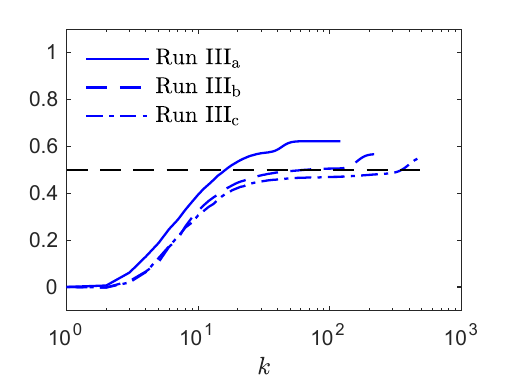}
\label{}
\end{subfigure}
\begin{subfigure}{0.32\textwidth}
\includegraphics[width=2.2 in]{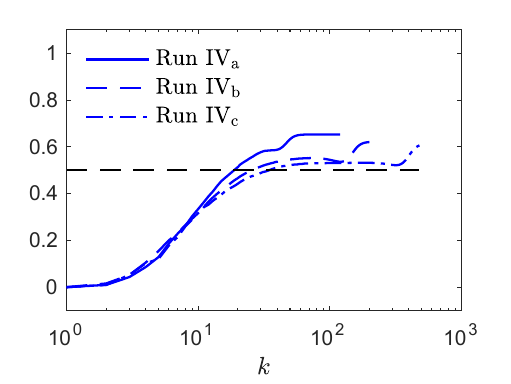}
\label{}
\end{subfigure}
\begin{subfigure}{0.32\textwidth}
\includegraphics[width=2.2 in]{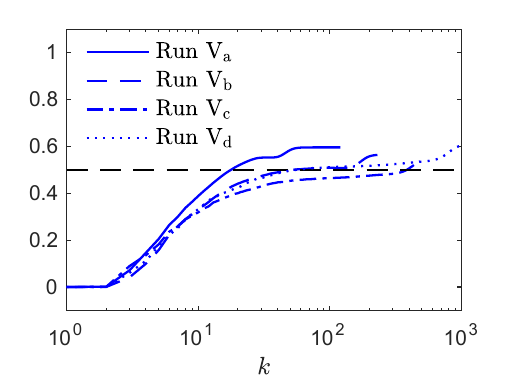}
\label{}
\end{subfigure}
\renewcommand{\figurename}{FIG.}
\caption{Plots showing convergence of $\langle\overline {S}_{ij}\overline{B}_i\overline{B}_j\rangle$ (blue lines) by increasing resolution from $256^3$ (solid line) to $512^3$ (dashed line) to $1{,}024^3$ (dot dashed line) to $2{,}048^3$ (dotted line, only applicable to Run V).}
\label{conversionConverg}
\end{figure*}

Figures \ref{totalfluxConverg} - \ref{conversionConverg} show plots of $\langle\overline \Pi_{\ell}\rangle$, $\langle\overline \Pi^u_{\ell}\rangle$,  $\langle\overline \Pi^b_{\ell}\rangle$,  and $\langle\overline {S}_{ij}\overline{B}_i\overline{B}_j\rangle$, respectively. Each Figure shows plots at different resolutions, indicating convergence of our results. The conversion range extent, $L\gg\ell>\ell_s$, does not keep increasing with resolution, but the the decoupled range, $\ell_s>\ell\gg\ell_d$, does. Moreover, $\langle\overline \Pi^u_{\ell}\rangle\approx\langle\overline \Pi^b_{\ell}\rangle$ over the decoupled range. 

\begin{figure*}
\centering
\begin{subfigure}{0.32\textwidth}
\includegraphics[width=2.2 in]{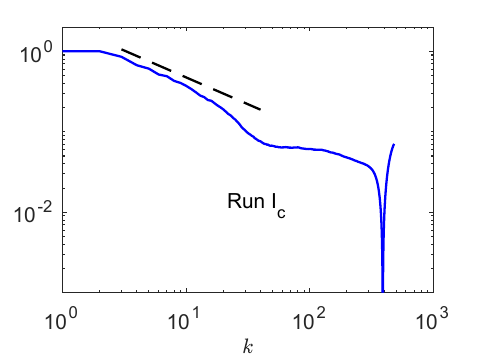}
\label{}
\end{subfigure}
\vspace{-0.10cm}
\begin{subfigure}{0.32\textwidth}
\includegraphics[width=2.2 in]{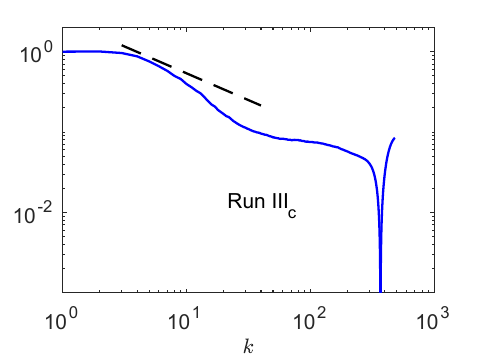}
\label{}
\end{subfigure}
\begin{subfigure}{0.32\textwidth}
\includegraphics[width=2.2 in]{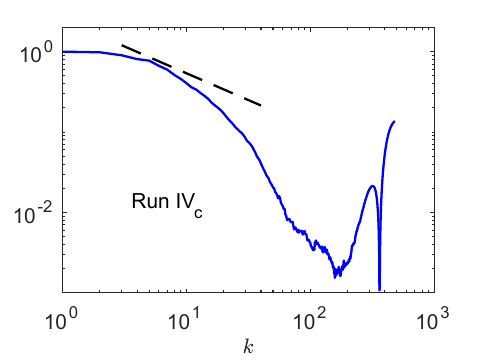}
\label{}
\end{subfigure}
\renewcommand{\figurename}{FIG.}
\caption{Plots of the relative residual conversion $\mR(k)/\mC_d = \langle \OL\bB_{\ell_d}\bdot\OL\bS_{\ell_d}\bdot\OL\bB_{\ell_d} -\OL\bB_{\ell}\bdot\OL\bS_{\ell}\bdot\OL\bB_{\ell} \rangle /\langle \OL\bB_{\ell_d}\bdot\OL\bS_{\ell_d}\bdot\OL\bB_{\ell_d} \rangle$  at the highest resolution from different Runs (see also Fig. 1 in the manuscript). The reference line with slope of -2/3 (black dashed line) is added, suggesting that KE-to-ME conversion saturates in a manner consistent with scale-locality \cite{AluieEyink10}.}
\label{fig:ScalingResidualConversion}
\end{figure*}

Figure \ref{fig:ScalingResidualConversion} shows the relative residual conversion 
\begin{eqnarray}
&\mR&(k)/\mC_d \nonumber\\
&=& \langle \OL\bB_{\ell_d}\bdot\OL\bS_{\ell_d}\bdot\OL\bB_{\ell_d} -\OL\bB_{\ell}\bdot\OL\bS_{\ell}\bdot\OL\bB_{\ell} \rangle /\langle \OL\bB_{\ell_d}\bdot\OL\bS_{\ell_d}\bdot\OL\bB_{\ell_d} \rangle~~~~\nonumber
\end{eqnarray}
from different Runs $\rm I_c$, $\rm III_c$, and $\rm IV_c$. See Fig. 1 in the manuscript for Runs $\rm II_c$ and $\rm V_d$. The reference line with slope of -2/3 (black dashed line) is added, suggesting that KE-to-ME conversion saturates in a manner consistent with scale-locality \cite{AluieEyink10}. Summarizing (non-formally) the scale-locality analysis of \cite{AluieEyink10}: the contribution to the strain $\OL\bS_\ell$ or current $\OL\bJ_\ell$ in the energy flux $\OL\Pi_\ell$ (or $\OL\Pi^u_\ell$ and $\OL\Pi^b_\ell$ separately) across scale $\ell$ from larger scales $L>\ell$ falls off as $(\ell/L)^{1-\sigma}$, where $\sigma$ is the scaling exponent of increments. Therefore, if $\sigma=1/3$ as in Goldreich-Sridhar's theory \cite{GoldreichSridhar95} (and Kolmogorov's 1941 theory) or if $\sigma=1/4$ as in Boldyrev's theory \cite{Boldyrev05}, the decay will be \emph{at least as fast as} $(\ell/L)^{2/3}$ or $(\ell/L)^{3/4}$, respectively. Moreover, the scaling $(\ell/L)^{1-\sigma}$ does not account for possible decorrelation effects, which can lead to even more rapid decay (see Fig. 1 and the associated discussion in \cite{EyinkAluie09,AluieEyink09}). Therefore, the decay rate is \emph{at least} $(\ell/L)^{1-\sigma}$ according to the formal analysis. Faster decay rates may arise due to decorrelation effects, but these cannot be guaranteed by the rigorous derivations in \cite{Eyink05,AluieEyink10}.

\begin{figure*}
\centering
\begin{subfigure}{0.32\textwidth}
\includegraphics[width=2.2 in]{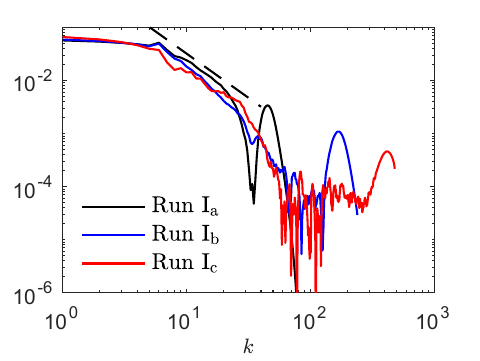}
\label{}
\end{subfigure}
\begin{subfigure}{0.32\textwidth}
\includegraphics[width=2.2 in]{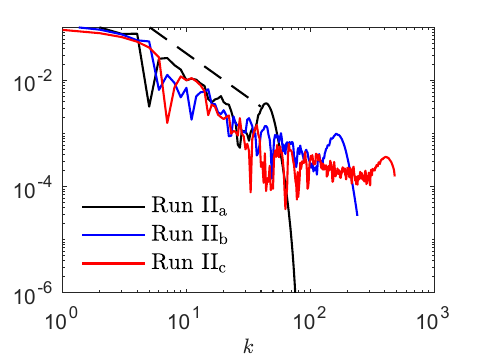}
\label{}
\end{subfigure}\\
\vspace{-0.10cm}
\begin{subfigure}{0.32\textwidth}
\includegraphics[width=2.2 in]{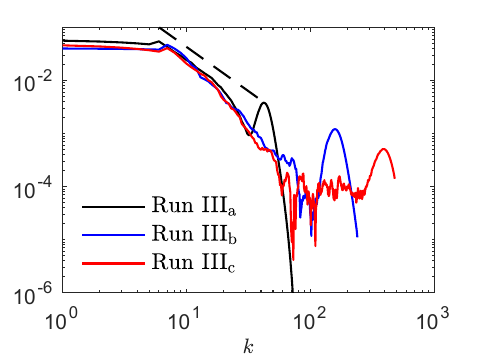}
\label{}
\end{subfigure}
\begin{subfigure}{0.32\textwidth}
\includegraphics[width=2.2 in]{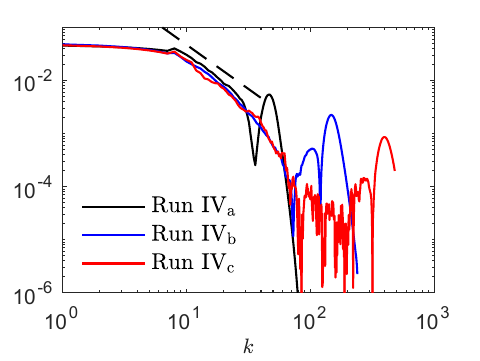}
\label{}
\end{subfigure}
\begin{subfigure}{0.32\textwidth}
\includegraphics[width=2.2 in]{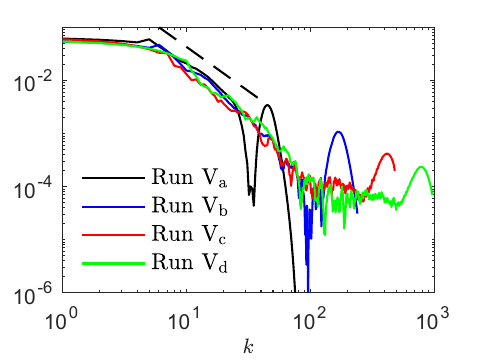}
\label{}
\end{subfigure}
\renewcommand{\figurename}{FIG.}
\caption{Plots of $\OL{C}^{ub}(k)$, the generalized filtering spectrum of $\langle {S}_{ij} {B}_i {B}_j\rangle$ defined in Eq. \eqref{eq:DefFilteringSpectrum} above.
The reference line (black dashed line) has a slope of -5/3. The five panels show $\OL{C}^{ub}(k)$ from each Run at different resolutions. In all cases, the conversion spectrum decays faster than $k^{-5/3}$, consistent with the residual conversion decaying faster than $k^{-2/3}$ in Fig. \ref{fig:ScalingResidualConversion}. Any decay faster than $k^{-1}$ within the inertial-inductive range is sufficient for
the saturation of $\langle\OL\bB_{\ell}\bdot\OL\bS_{\ell}\bdot\OL\bB_{\ell} \rangle$ observed in Fig. \ref{conversionConverg}.}
\label{conversionSpec}
\end{figure*}

Figure \ref{conversionSpec} shows the generalized ``filtering spectrum'' of conversion, defined as
\be
\OL{C}^{ub}(k) \equiv \frac{d}{dk}\mC(\ell),
\lb{eq:DefFilteringSpectrum}\ee
where $k_\ell=2\pi/\ell$. The filtering spectrum was introduced recently in \cite{SadekAluie18} and is consistent with the traditional Fourier spectrum for quadratic quantities such as energy \cite{Frisch95}. One of its main advantages lies in calculating spectra of non-quadratic quantities, such as $\bB\bdot\bS_{}\bdot\bB_{}$, in a manner consistent with the chosen scale decomposition rather than having to treat the quantities as quadratic.
The conversion spectrum seems to follow $\OL{C}^{ub}(k)\sim k^{-\beta}$ with $\beta \ge 5/3$, consistent with the residual conversion decaying faster than $k^{-2/3}$ in Fig. \ref{fig:ScalingResidualConversion}.
Since $\beta >1$, its integral, which is the cumulative mean KE-to-ME conversion at all scales $>\ell$, has to saturate (converge) in the limit $\ell\to\ell_d\to0$, consistent with the saturation of $\langle\OL\bB_{\ell}\bdot\OL\bS_{\ell}\bdot\OL\bB_{\ell} \rangle$ observed in Fig. \ref{conversionConverg}.

\begin{figure*}
\centering
\begin{subfigure}{0.45\textwidth}
\includegraphics[width=2.2 in]{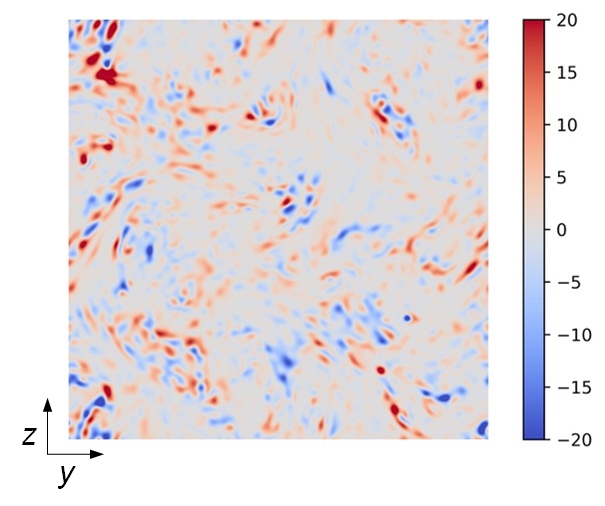}
\end{subfigure}
\begin{subfigure}{0.45\textwidth}
\includegraphics[width=2.2 in]{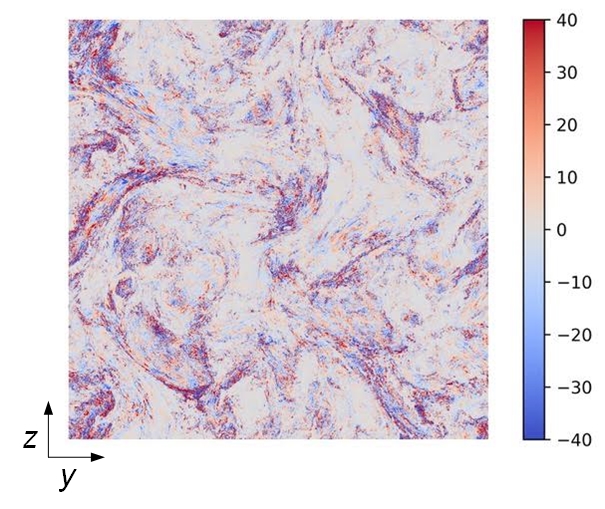}
\end{subfigure}
\\
\centering
\hspace{-1cm}
\begin{subfigure}{0.45\textwidth}
\includegraphics[width=2.2 in]{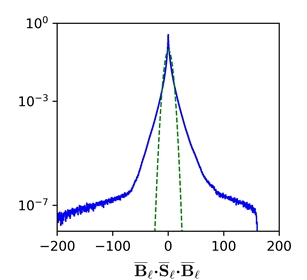}
\end{subfigure}
\begin{subfigure}{0.45\textwidth}
\includegraphics[width=2.2 in]{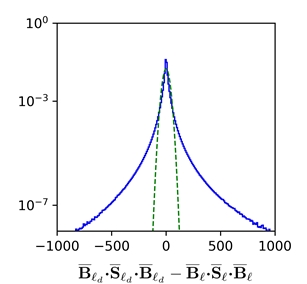}
\end{subfigure}
\renewcommand{\figurename}{FIG.}
\caption{For scale $\ell=2\pi/30$ ($k=30$) from Run $\rm I_c$ at one instant in time (see Fig. 1 in the Letter or Fig. \ref{conversionConverg} above): top two panels show a 2D slice from the 3D domain of pointwise conversion at large scales, $\OL\bB_{\ell}\bdot\OL\bS_{\ell}\bdot\OL\bB_{\ell} (\bx)$ (top left), and small scales, $\OL\bB_{\ell_d}\bdot\OL\bS_{\ell_d}\bdot\OL\bB_{\ell_d} (\bx) -\OL\bB_{\ell}\bdot\OL\bS_{\ell}\bdot\OL\bB_{\ell} (\bx)$  (top right).  
Bottom two panels show probability density function of conversion as a function of $\bx$ at large scales (bottom left) and small scales (bottom right). The large-scale distribution has mean of 0.46 and variance of 18.63. The small-scale distribution has mean of 0.05 and variance of 513.05. Quantities are normalized by energy injection rate $\epsilon^{\inj}$. Unnormalized Gaussians (green dashed lines) are added to both plots.}
\label{mhdvis}
\end{figure*}

Figure \ref{mhdvis} (similar to Fig. 2 in the main Letter) shows how the decoupling between the mean ME and KE budgets arises from decorrelation effects at small scales. For an intermediate scale $\ell=2\pi/30$ from Run $\rm I_c$ in the absence of a uniform background $\bB_0$ field, Fig. \ref{mhdvis} shows how magnetic field-line stretching, which is concentrated in magnetic filaments, is an order of magnitude more intense at scales smaller than $\ell=2\pi/30$ compared to larger scales. Yet, the small-scale contribution fluctuates vigorously in sign, yielding a mere $10\%$ to the space average. 

\begin{figure*}
\centering
\begin{subfigure}{0.32\textwidth}
\includegraphics[width=2.2 in]{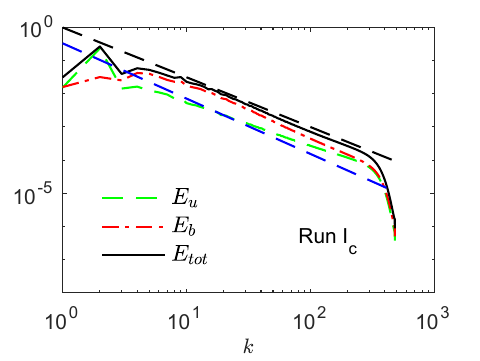}
\end{subfigure}
\begin{subfigure}{0.32\textwidth}
\includegraphics[width=2.2 in]{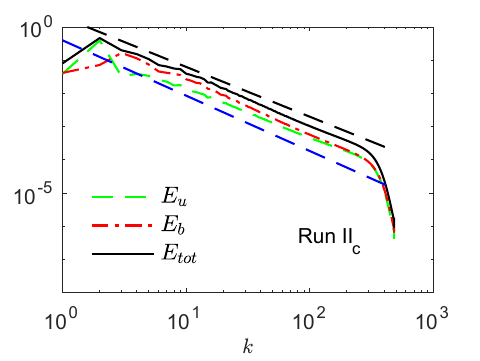}
\end{subfigure}
\\
\vspace{-0.1cm}
\begin{subfigure}{0.32\textwidth}
\includegraphics[width=2.2 in]{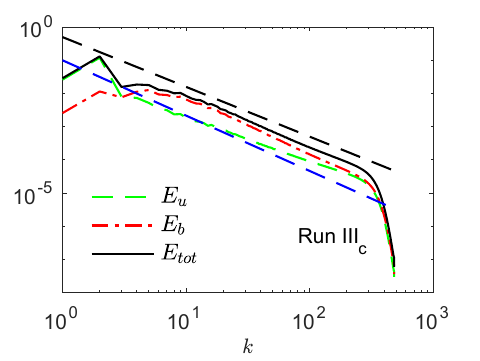}
\end{subfigure}
\begin{subfigure}{0.32\textwidth}
\includegraphics[width=2.2 in]{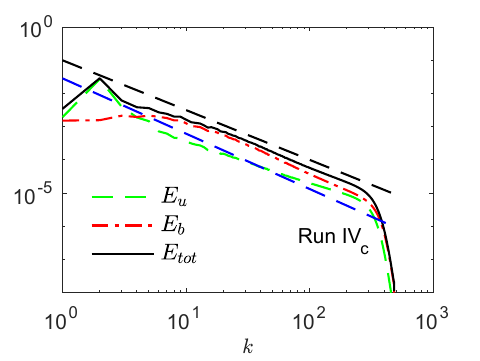}
\end{subfigure}
\begin{subfigure}{0.32\textwidth}
\includegraphics[width=2.2 in]{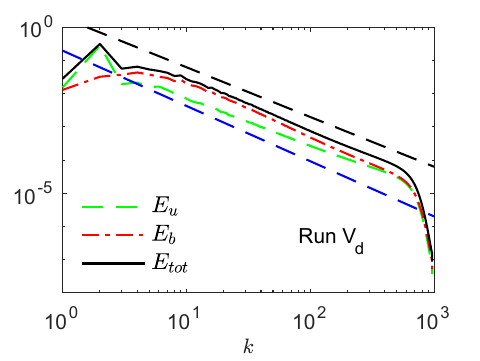}
\end{subfigure}
\renewcommand{\figurename}{FIG.}
\caption{Kinetic, magnetic, and total energy spectra, $E_u$, $E_b$, and $E_{tot}$, respectively. We show two reference lines with slopes -3/2 (black dashed line) and -5/3 (blue dashed line). We don't observe an obvious correlation between the wavenumber at which $E_u$ and $E_b$ cross and that of decoupling. 
}
\label{mhdspec}
\end{figure*}

\begin{figure*}
\centering
\begin{subfigure}{0.32\textwidth}
\includegraphics[width=2.2 in]{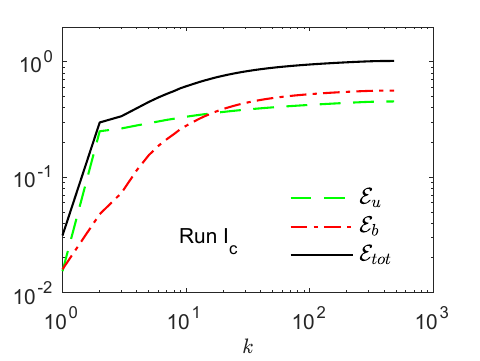}
\end{subfigure}
\begin{subfigure}{0.32\textwidth}
\includegraphics[width=2.2 in]{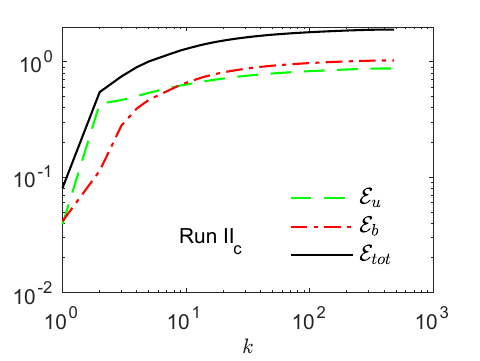}
\end{subfigure}
\\
\vspace{-0.1cm}
\begin{subfigure}{0.32\textwidth}
\includegraphics[width=2.2 in]{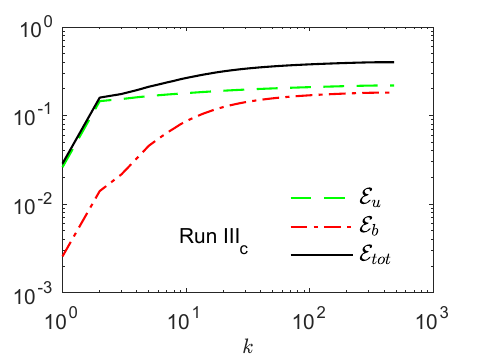}
\end{subfigure}
\begin{subfigure}{0.32\textwidth}
\includegraphics[width=2.2 in]{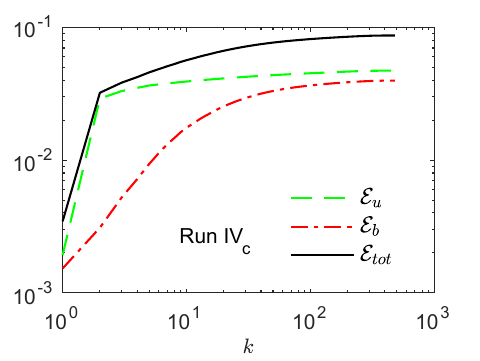}
\end{subfigure}
\begin{subfigure}{0.32\textwidth}
\includegraphics[width=2.2 in]{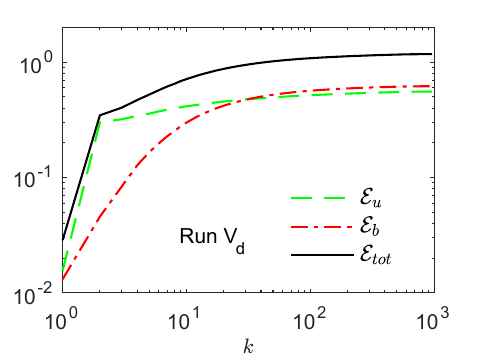}
\end{subfigure}
\renewcommand{\figurename}{FIG.}
\caption{Cumulative kinetic, magnetic, and total energy spectra, ${\mathcal E}_{u,b,tot}(K)= \sum_{k=0}^{k=K}E_{u,b,tot}(k)$, which quantify energy at all wavenumbers $\le K$. We don't observe an obvious correlation between the wavenumber at which ${\mathcal E}_{u}$ and ${\mathcal E}_{b}$ cross and that of decoupling.}
\label{fig:mhdCumspec}
\end{figure*}

Figure \ref{mhdspec} shows KE, ME, and total energy spectra from Runs I to V at the highest resolution in this study. The scale at which the mean KE and ME budgets decouple (scale at which $\langle\OL\Pi^u\rangle$ and $\langle\OL\Pi^b\rangle$ reach equipartition) is $30\lesssim k_s \lesssim 100$ (Fig. 1 in Letter). However, $E_u(k)$ and $E_b(k)$ do not reach equipartition anywhere within the inertial-inductive range, as is well-known. Note that the wavenumber $k_{crs}$ at which the spectra cross, $E_u(k)=E_b(k)$, is within $2\le k_{crs} \le 4$ in all five simulations and does not show a correlation with $k_s$. 
We have also compared the cumulative spectra, ${\mathcal E}_{u,b}(K)\equiv \sum_{k=0}^{k=K}E_{u,b}(k)$ in Fig. \ref{fig:mhdCumspec} and found no correlation between the wavenumber at which they cross and that of decoupling.

%-------------------------------------------------------------------------------------------------------------------------------------------------------------------------------------------
%-------------------------------------------------------------------------------------------------------------------------------------------------------------------------------------------

\clearpage
\subsection{Non-colliding waves}
The role of waves is further illustrated via two simulations of non-colliding Alfv\'en waves, (i) a spatially localized Alfv\'en wavepacket (ii) a monochromatic (single mode) Alfv\'en wave. The main purpose is pedagogical: these well-known examples help demonstrate the behavior of $\langle\OL\bB_{\ell}\bdot\OL\bS_{\ell}\bdot\OL\bB_{\ell} \rangle$ as a function of $\ell$. In these simple solutions, the nonlinearity is identically zero and different modes cannot exchange energy. Energy is only converted between KE and ME forms \emph{at the same} $k$. Note that this is an unsteady unforced problem to which our eqs. (7)-(10) in the Letter do not apply.

\subsubsection{Simulation Details}
\begin{figure*}
\centering
\includegraphics[width=2.4 in]{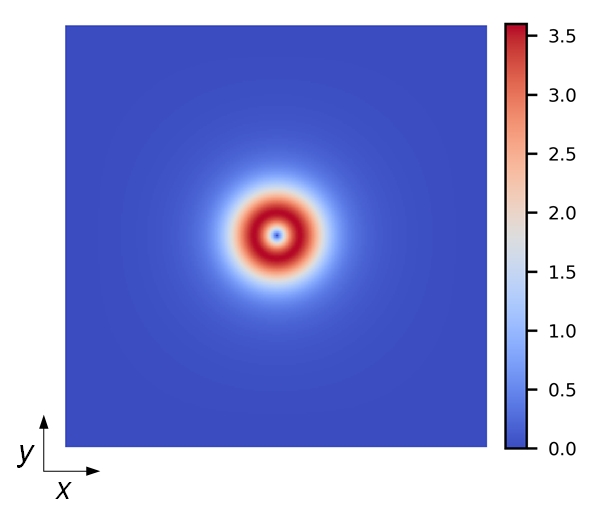} 
\renewcommand{\figurename}{FIG.}
\caption{Initial condition for the wavepacket simulation. The plot shows $|\bu|$ in a 2D slice at $z=\pi$. }
\label{fig:wavepack_init}
\end{figure*}

The wavepacket simulation is solved in a $256^3$ periodic box $\mathbb{T}^3=[0, 2\pi)^3$ with an external magnetic field $|\textbf{B}_0|=10$ in the $z$ direction and with $\nu_h=\eta_h=5\times 10^{-10}$. The simulation is initialized within a spatially localized region (in all three directions). Fig. \ref{fig:wavepack_init} shows a 2D slice. This results in two counter-propagating wavepackets. We restrict our analysis to times before the two packets collide.

The monochromatic wave simulation is solved in a $128^3$ periodic box $\mathbb{T}^3=[0, 2\pi)^3$ with an external magnetic field $|\textbf{B}_0|=10$ in the $z$ direction and with $\nu_h=\eta_h=5\times 10^{-10}$. The velocity is initialized in Fourier space:
\begin{eqnarray}
\hat{u}_x(\bk)&=&\begin{cases}
    1, & \text{if $\bk=(k_x,k_y,k_z)=(0,0,\pm8)$}.\\
    0, & \text{otherwise}.\\
  \end{cases}\\
\hat{u}_y(\bk) &=& \hat{u}_z(\bk) =0.
\lb{app_eq:MonochromaticWave}\end{eqnarray}
This generates a sinusoidal wave at scale $\ell=2\pi/8$ ($k=8$) that is uniform in the $x-y$ directions and propagates in $z$.

\subsubsection{Wavepacket}

\begin{figure*}
\centering
\begin{subfigure}{0.45\textwidth}
\includegraphics[width=2.4 in]{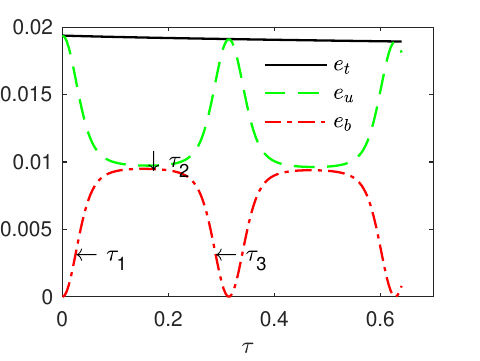} 
\end{subfigure}
\begin{subfigure}{0.45\textwidth}
\includegraphics[width=2.4 in]{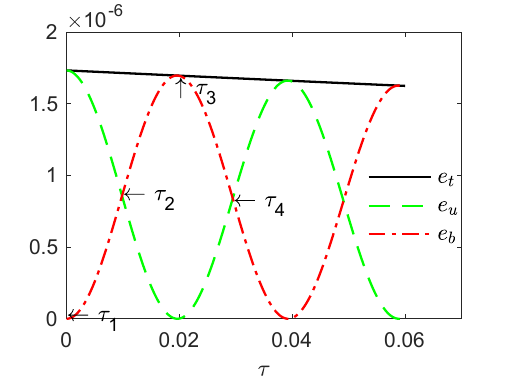}
\end{subfigure}
\renewcommand{\figurename}{FIG.}
\caption{Time series of total energy $e_t$, kinetic energy $e_u$, and magnetic energy $e_b$ for the wavepacket (left), and the monochromatic wave (right). Times labelled $\tau_1$, $\tau_2$, $\tau_3$, and $\tau_4$ are those at which we analyze the energy budgets in Figs. \ref{vortexflux}-\ref{waveflux} below.
}
\label{vortex_wave_energy}
\end{figure*}

\begin{figure*}
\centering
\begin{subfigure}{0.32\textwidth}
\includegraphics[width=2.3 in]{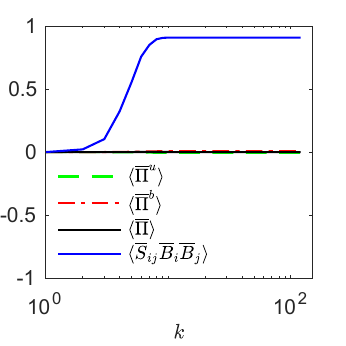}
\end{subfigure}
\begin{subfigure}{0.32\textwidth}
\includegraphics[width=2.3 in]{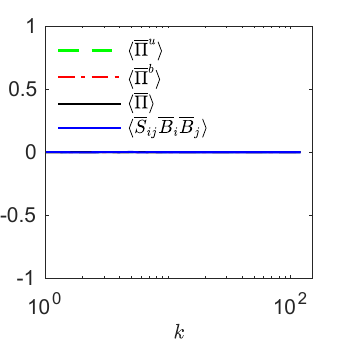}
\end{subfigure}
\begin{subfigure}{0.32\textwidth}
\includegraphics[width=2.3 in]{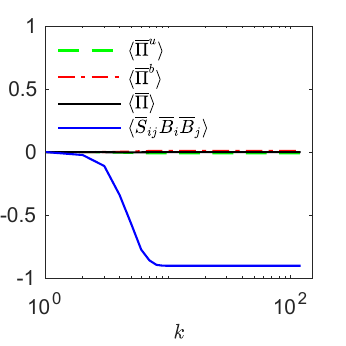}
\end{subfigure}
\begin{subfigure}{0.32\textwidth}
\includegraphics[width= 2.3 in]{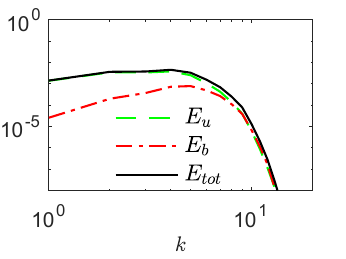}
\end{subfigure}
\begin{subfigure}{0.32\textwidth}
\includegraphics[width= 2.3 in]{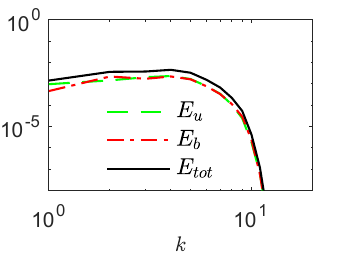}
\end{subfigure}
\begin{subfigure}{0.32\textwidth}
\includegraphics[width= 2.3 in]{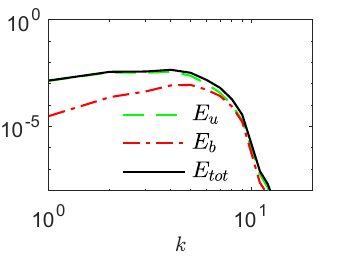}
\end{subfigure}
\renewcommand{\figurename}{FIG.}
\caption{Wavepacket: top three panels show 
$\langle\overline \Pi^u\rangle$, $\langle\overline \Pi^b\rangle$, $\langle\overline \Pi\rangle$ and $\langle\overline {S}_{ij}\overline{B}_i\overline{B}_j\rangle$ as a function of $k\equiv2\pi/\ell$ from the wavepacket example at $\tau_1$ (left), $\tau_2$ (middle), and $\tau_3$ (right). 
Bottom three panels show energy spectra at the corresponding times.
}
\label{vortexflux}
\end{figure*}

Fig. \ref{vortex_wave_energy} shows the time series of total energy $e_t$, kinetic energy $e_u$, and magnetic energy $e_b$.
In Fig. \ref{vortexflux}, we analyze the coarse-grained energy budget and plot the energy spectra 
at three different times $\tau_1$, $\tau_2$, and $\tau_3$ labeled in Fig. \ref{vortex_wave_energy}.
Figure \ref{vortexflux} shows the energy cascade terms and conversion at these three times. The cascade terms $\langle\overline \Pi^u_\ell\rangle=\langle\overline \Pi^b_\ell\rangle = \langle\overline \Pi_\ell\rangle=0$ at all scales and for all times, as expected.

From the spectra in Fig \ref{vortexflux}, we see that most of the energy lies in the range $k\in[2,8]$. The nonlinearity in this flow is identically zero and different modes cannot exchange energy. Energy is only converted between KE and ME forms {at the same} $k$. 
At $\tau_1$, when $e_b$ is increasing and $e_u$ is decreasing, we see that mean KE-to-ME conversion $\mC(\ell)\ge0$, increasing over the range $k\in[2,8]$, within the same band of scales populated by the wavepacket.
Similarly, at time $\tau_3$, when $e_b$ is decreasing and $e_u$ is increasing, $\mC(\ell)\le0$, decreasing over the same range $k\in[2,8]$.
At time $\tau_2$,  when $\partial_t e_b = \partial_t e_u=0$, we have $\mC(\ell)=0$ at all scales.

\subsubsection{Monochromatic Wave}
Fig. \ref{vortex_wave_energy} shows the time series of total energy $e_t$, kinetic energy $e_u$, and magnetic energy $e_b$ for the monochromatic wave. We analyze the coarse-grained energy budget at four different times $\tau_1$, $\tau_2$, $\tau_3$, and $\tau_4$ labeled in Fig. \ref{vortex_wave_energy}.

\begin{figure*}
\centering
\begin{subfigure}{0.45\textwidth}
\includegraphics[height=1.7 in,width=2.5 in]{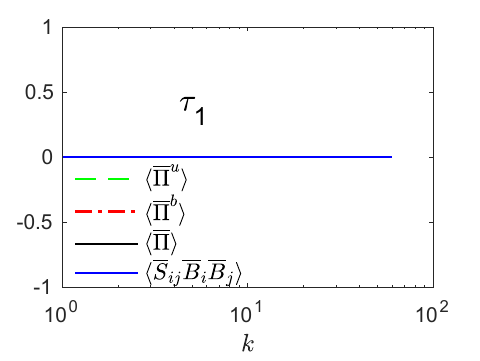}
\end{subfigure}
\begin{subfigure}{0.45\textwidth}
\includegraphics[height=1.7 in,width=2.5 in]{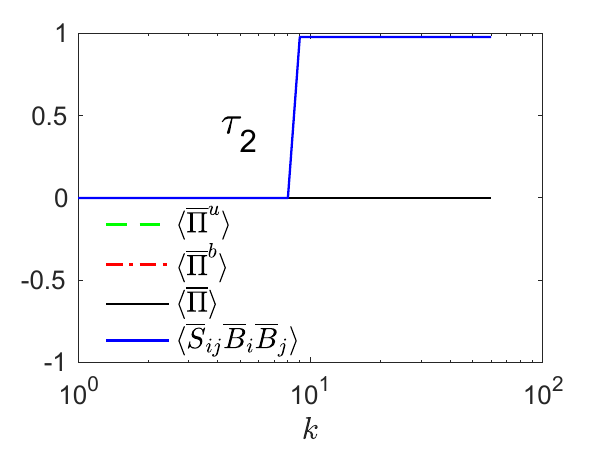}
\end{subfigure}\\
\begin{subfigure}{0.45\textwidth}
\includegraphics[height=1.7 in,width=2.5 in]{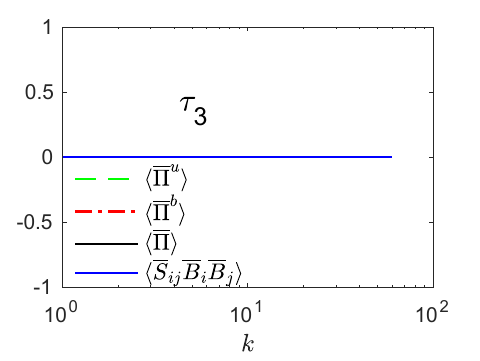}
\end{subfigure}
\begin{subfigure}{0.45\textwidth}
\includegraphics[height=1.7 in,width=2.5 in]{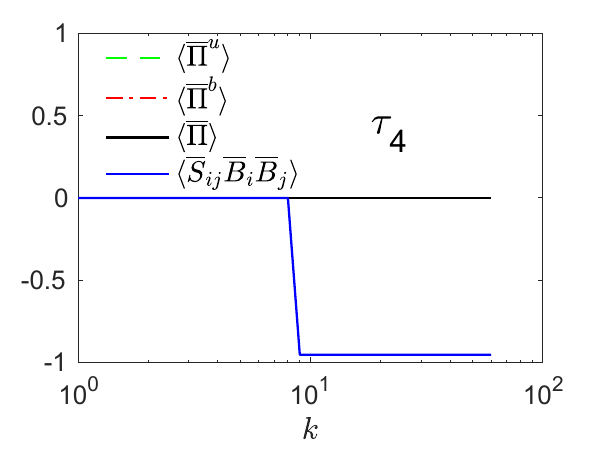}
\end{subfigure}
\renewcommand{\figurename}{FIG.}
\caption{Plots of $\langle\overline \Pi^u\rangle$, $\langle\overline \Pi^b\rangle$, $\langle\overline \Pi\rangle$ and $\langle\overline {S}_{ij}\overline{B}_i\overline{B}_j\rangle$ as a function of $k\equiv2\pi/\ell$ from the monochromatic wave example at $\tau_1$ (top left), $\tau_2$ (top right), $\tau_3$ (bottom left), and $\tau_4$ (bottom right). 
}
\label{waveflux}
\end{figure*}

Figure \ref{waveflux} shows the cascade terms $\langle\overline \Pi^u_\ell\rangle=\langle\overline \Pi^b_\ell\rangle = \langle\overline \Pi_\ell\rangle=0$ at all scales and for all times, as expected.
At $\tau_2$, when $e_b$ is increasing and $e_u$ is decreasing, $\mC(\ell)$ shows a discontinuous jump from zero to a positive value at $k=8$, the wave's wavenumber. Similarly, at time $\tau_3$ when $e_b$ is decreasing and $e_u$ is increasing, $\mC(\ell)$ shows a discontinuous jump from zero to a negative value at $k=8$. At times $\tau_1$ and $\tau_3$, when $\partial_t e_b = \partial_t e_u=0$, we have $\mC(\ell)=0$ at all scales.

\bibliography{BianAluie_MHD}

%merlin.mbs apsrev4-1.bst 2010-07-25 4.21a (PWD, AO, DPC) hacked
%Control: key (0)
%Control: author (8) initials jnrlst
%Control: editor formatted (1) identically to author
%Control: production of article title (-1) disabled
%Control: page (0) single
%Control: year (1) truncated
%Control: production of eprint (0) enabled
\begin{thebibliography}{64}%
\makeatletter
\providecommand \@ifxundefined [1]{%
 \@ifx{#1\undefined}
}%
\providecommand \@ifnum [1]{%
 \ifnum #1\expandafter \@firstoftwo
 \else \expandafter \@secondoftwo
 \fi
}%
\providecommand \@ifx [1]{%
 \ifx #1\expandafter \@firstoftwo
 \else \expandafter \@secondoftwo
 \fi
}%
\providecommand \natexlab [1]{#1}%
\providecommand \enquote  [1]{``#1''}%
\providecommand \bibnamefont  [1]{#1}%
\providecommand \bibfnamefont [1]{#1}%
\providecommand \citenamefont [1]{#1}%
\providecommand \href@noop [0]{\@secondoftwo}%
\providecommand \href [0]{\begingroup \@sanitize@url \@href}%
\providecommand \@href[1]{\@@startlink{#1}\@@href}%
\providecommand \@@href[1]{\endgroup#1\@@endlink}%
\providecommand \@sanitize@url [0]{\catcode `\\12\catcode `\$12\catcode
  `\&12\catcode `\#12\catcode `\^12\catcode `\_12\catcode `\%12\relax}%
\providecommand \@@startlink[1]{}%
\providecommand \@@endlink[0]{}%
\providecommand \url  [0]{\begingroup\@sanitize@url \@url }%
\providecommand \@url [1]{\endgroup\@href {#1}{\urlprefix }}%
\providecommand \urlprefix  [0]{URL }%
\providecommand \Eprint [0]{\href }%
\providecommand \doibase [0]{http://dx.doi.org/}%
\providecommand \selectlanguage [0]{\@gobble}%
\providecommand \bibinfo  [0]{\@secondoftwo}%
\providecommand \bibfield  [0]{\@secondoftwo}%
\providecommand \translation [1]{[#1]}%
\providecommand \BibitemOpen [0]{}%
\providecommand \bibitemStop [0]{}%
\providecommand \bibitemNoStop [0]{.\EOS\space}%
\providecommand \EOS [0]{\spacefactor3000\relax}%
\providecommand \BibitemShut  [1]{\csname bibitem#1\endcsname}%
\let\auto@bib@innerbib\@empty
%</preamble>
\bibitem [{\citenamefont {{Iroshnikov}}(1964)}]{Iroshnikov64}%
  \BibitemOpen
  \bibfield  {author} {\bibinfo {author} {\bibfnamefont {P.~S.}\ \bibnamefont
  {{Iroshnikov}}},\ }\href@noop {} {\bibfield  {journal} {\bibinfo  {journal}
  {Sov. Astron.}\ }\textbf {\bibinfo {volume} {7}},\ \bibinfo {pages} {566}
  (\bibinfo {year} {1964})}\BibitemShut {NoStop}%
\bibitem [{\citenamefont {{Kraichnan}}(1965)}]{Kraichnan65}%
  \BibitemOpen
  \bibfield  {author} {\bibinfo {author} {\bibfnamefont {R.~H.}\ \bibnamefont
  {{Kraichnan}}},\ }\href {\doibase 10.1063/1.1761412} {\bibfield  {journal}
  {\bibinfo  {journal} {Phys. Fluids}\ }\textbf {\bibinfo {volume} {8}},\
  \bibinfo {pages} {1385} (\bibinfo {year} {1965})}\BibitemShut {NoStop}%
\bibitem [{\citenamefont {{Goldreich}}\ and\ \citenamefont
  {{Sridhar}}(1995)}]{GoldreichSridhar95}%
  \BibitemOpen
  \bibfield  {author} {\bibinfo {author} {\bibfnamefont {P.}~\bibnamefont
  {{Goldreich}}}\ and\ \bibinfo {author} {\bibfnamefont {S.}~\bibnamefont
  {{Sridhar}}},\ }\href {\doibase 10.1086/175121} {\bibfield  {journal}
  {\bibinfo  {journal} {Astrophys. J.}\ }\textbf {\bibinfo {volume} {438}},\
  \bibinfo {pages} {763} (\bibinfo {year} {1995})}\BibitemShut {NoStop}%
\bibitem [{\citenamefont {{Boldyrev}}(2005)}]{Boldyrev05}%
  \BibitemOpen
  \bibfield  {author} {\bibinfo {author} {\bibfnamefont {S.}~\bibnamefont
  {{Boldyrev}}},\ }\href {\doibase 10.1086/431649} {\bibfield  {journal}
  {\bibinfo  {journal} {Astrophys. J.}\ }\textbf {\bibinfo {volume} {626}},\
  \bibinfo {pages} {L37} (\bibinfo {year} {2005})}\BibitemShut {NoStop}%
\bibitem [{\citenamefont {Boldyrev}(2006)}]{Boldyrev06}%
  \BibitemOpen
  \bibfield  {author} {\bibinfo {author} {\bibfnamefont {S.}~\bibnamefont
  {Boldyrev}},\ }\href@noop {} {\bibfield  {journal} {\bibinfo  {journal}
  {Physical Review Letters}\ }\textbf {\bibinfo {volume} {96}},\ \bibinfo
  {pages} {742} (\bibinfo {year} {2006})}\BibitemShut {NoStop}%
\bibitem [{\citenamefont {Aluie}\ and\ \citenamefont
  {Eyink}(2010)}]{AluieEyink10}%
  \BibitemOpen
  \bibfield  {author} {\bibinfo {author} {\bibfnamefont {H.}~\bibnamefont
  {Aluie}}\ and\ \bibinfo {author} {\bibfnamefont {G.~L.}\ \bibnamefont
  {Eyink}},\ }\href@noop {} {\bibfield  {journal} {\bibinfo  {journal}
  {Physical review letters}\ }\textbf {\bibinfo {volume} {104}},\ \bibinfo
  {pages} {081101} (\bibinfo {year} {2010})}\BibitemShut {NoStop}%
\bibitem [{\citenamefont {Eyink}(2005)}]{Eyink05}%
  \BibitemOpen
  \bibfield  {author} {\bibinfo {author} {\bibfnamefont {G.~L.}\ \bibnamefont
  {Eyink}},\ }\href@noop {} {\bibfield  {journal} {\bibinfo  {journal} {Physica
  D: Nonlinear Phenomena}\ }\textbf {\bibinfo {volume} {207}},\ \bibinfo
  {pages} {91} (\bibinfo {year} {2005})}\BibitemShut {NoStop}%
\bibitem [{\citenamefont {Aluie}(2009)}]{AluieThesis}%
  \BibitemOpen
  \bibfield  {author} {\bibinfo {author} {\bibfnamefont {H.}~\bibnamefont
  {Aluie}},\ }\href@noop {} {Ph.D. thesis},\ \bibinfo  {school} {The Johns
  Hopkins University}, \bibinfo {address} {Baltimore} (\bibinfo {year}
  {2009}),\ \bibinfo {note} {pp. 161--166.}\BibitemShut {Stop}%
\bibitem [{\citenamefont {{Aluie}}(2010)}]{Aluie10}%
  \BibitemOpen
  \bibfield  {author} {\bibinfo {author} {\bibfnamefont {H.}~\bibnamefont
  {{Aluie}}},\ }\href@noop {} {\bibfield  {journal} {\bibinfo  {journal} {AGU
  Fall Meeting Abstracts}\ ,\ \bibinfo {eid} {SH41B-1785}} (\bibinfo {year}
  {2010})}\BibitemShut {NoStop}%
\bibitem [{\citenamefont {Shakura}\ and\ \citenamefont
  {Sunyaev}(1973)}]{ShakuraSunyaev1973}%
  \BibitemOpen
  \bibfield  {author} {\bibinfo {author} {\bibfnamefont {N.~I.}\ \bibnamefont
  {Shakura}}\ and\ \bibinfo {author} {\bibfnamefont {R.~A.}\ \bibnamefont
  {Sunyaev}},\ }\href@noop {} {\bibfield  {journal} {\bibinfo  {journal}
  {Astronomy and Astrophysics}\ }\textbf {\bibinfo {volume} {24}},\ \bibinfo
  {pages} {337} (\bibinfo {year} {1973})}\BibitemShut {NoStop}%
\bibitem [{\citenamefont {Blandford}\ and\ \citenamefont
  {Payne}(1982)}]{BlandfordPayne1982}%
  \BibitemOpen
  \bibfield  {author} {\bibinfo {author} {\bibfnamefont {R.~D.}\ \bibnamefont
  {Blandford}}\ and\ \bibinfo {author} {\bibfnamefont {D.~G.}\ \bibnamefont
  {Payne}},\ }\href@noop {} {\bibfield  {journal} {\bibinfo  {journal} {Monthly
  Notices of the Royal Astronomical Society}\ }\textbf {\bibinfo {volume}
  {199}},\ \bibinfo {pages} {883} (\bibinfo {year} {1982})}\BibitemShut
  {NoStop}%
\bibitem [{\citenamefont {Hawley}\ \emph {et~al.}(1995)\citenamefont {Hawley},
  \citenamefont {Gammie},\ and\ \citenamefont {Balbus}}]{Hawleyetal95}%
  \BibitemOpen
  \bibfield  {author} {\bibinfo {author} {\bibfnamefont {J.~F.}\ \bibnamefont
  {Hawley}}, \bibinfo {author} {\bibfnamefont {C.~F.}\ \bibnamefont {Gammie}},
  \ and\ \bibinfo {author} {\bibfnamefont {S.~A.}\ \bibnamefont {Balbus}},\
  }\href@noop {} {\bibfield  {journal} {\bibinfo  {journal} {Astrophysical
  Journal}\ }\textbf {\bibinfo {volume} {440}},\ \bibinfo {pages} {742}
  (\bibinfo {year} {1995})}\BibitemShut {NoStop}%
\bibitem [{\citenamefont {Guan}\ and\ \citenamefont
  {Gammie}(2009)}]{GuanGammie09}%
  \BibitemOpen
  \bibfield  {author} {\bibinfo {author} {\bibfnamefont {X.}~\bibnamefont
  {Guan}}\ and\ \bibinfo {author} {\bibfnamefont {C.~F.}\ \bibnamefont
  {Gammie}},\ }\href@noop {} {\bibfield  {journal} {\bibinfo  {journal} {The
  Astrophysical Journal Letters}\ }\textbf {\bibinfo {volume} {697}},\ \bibinfo
  {pages} {1901} (\bibinfo {year} {2009})}\BibitemShut {NoStop}%
\bibitem [{\citenamefont {Lesur}\ and\ \citenamefont
  {Longaretti}(2009)}]{LesurLongaretti09}%
  \BibitemOpen
  \bibfield  {author} {\bibinfo {author} {\bibfnamefont {G.}~\bibnamefont
  {Lesur}}\ and\ \bibinfo {author} {\bibfnamefont {P.~Y.}\ \bibnamefont
  {Longaretti}},\ }\href@noop {} {\bibfield  {journal} {\bibinfo  {journal}
  {Astronomy and Astrophysics}\ }\textbf {\bibinfo {volume} {504}},\ \bibinfo
  {pages} {309} (\bibinfo {year} {2009})}\BibitemShut {NoStop}%
\bibitem [{\citenamefont {Fromang}\ and\ \citenamefont
  {Stone}(2009)}]{FromangStone09}%
  \BibitemOpen
  \bibfield  {author} {\bibinfo {author} {\bibfnamefont {S.}~\bibnamefont
  {Fromang}}\ and\ \bibinfo {author} {\bibfnamefont {J.~M.}\ \bibnamefont
  {Stone}},\ }\href@noop {} {\bibfield  {journal} {\bibinfo  {journal}
  {Astronomy {\&} Astrophysics}\ }\textbf {\bibinfo {volume} {507}},\ \bibinfo
  {pages} {19} (\bibinfo {year} {2009})}\BibitemShut {NoStop}%
\bibitem [{\citenamefont {Cao}(2011)}]{Cao11}%
  \BibitemOpen
  \bibfield  {author} {\bibinfo {author} {\bibfnamefont {X.}~\bibnamefont
  {Cao}},\ }\href@noop {} {\bibfield  {journal} {\bibinfo  {journal} {The
  Astrophysical Journal Letters}\ }\textbf {\bibinfo {volume} {737}},\ \bibinfo
  {pages} {94} (\bibinfo {year} {2011})}\BibitemShut {NoStop}%
\bibitem [{\citenamefont {Aluie}(2017)}]{Aluie17}%
  \BibitemOpen
  \bibfield  {author} {\bibinfo {author} {\bibfnamefont {H.}~\bibnamefont
  {Aluie}},\ }\href@noop {} {\bibfield  {journal} {\bibinfo  {journal} {New J.
  Phys.}\ }\textbf {\bibinfo {volume} {19}},\ \bibinfo {pages} {025008}
  (\bibinfo {year} {2017})}\BibitemShut {NoStop}%
\bibitem [{\citenamefont {{Moffatt}}(1978)}]{Moffatt78}%
  \BibitemOpen
  \bibfield  {author} {\bibinfo {author} {\bibfnamefont {H.~K.}\ \bibnamefont
  {{Moffatt}}},\ }\href@noop {} {\emph {\bibinfo {title} {{Magnetic field
  generation in electrically conducting fluids}}}}\ (\bibinfo  {publisher}
  {Cambridge University Press, Cambridge, England},\ \bibinfo {year}
  {1978})\BibitemShut {NoStop}%
\bibitem [{\citenamefont {{Krause}}\ and\ \citenamefont
  {{R\"{a}dler}}(1980)}]{KrauseRaedler80}%
  \BibitemOpen
  \bibfield  {author} {\bibinfo {author} {\bibfnamefont {F.}~\bibnamefont
  {{Krause}}}\ and\ \bibinfo {author} {\bibfnamefont {K.~H.}\ \bibnamefont
  {{R\"{a}dler}}},\ }\href@noop {} {\emph {\bibinfo {title} {{Mean-Field
  Magnetohydrodynamics and Dynamo Theory.}}}}\ (\bibinfo  {publisher} {Pergamon
  Press, New York},\ \bibinfo {year} {1980})\BibitemShut {NoStop}%
\bibitem [{\citenamefont {Eyink}(2018)}]{Eyink18}%
  \BibitemOpen
  \bibfield  {author} {\bibinfo {author} {\bibfnamefont {G.~L.}\ \bibnamefont
  {Eyink}},\ }\href@noop {} {\bibfield  {journal} {\bibinfo  {journal}
  {arXiv.org}\ } (\bibinfo {year} {2018})},\ \Eprint
  {http://arxiv.org/abs/1803.02223v2} {1803.02223v2} \BibitemShut {NoStop}%
\bibitem [{\citenamefont {Zhao}\ and\ \citenamefont
  {Aluie}(2018)}]{ZhaoAluie18}%
  \BibitemOpen
  \bibfield  {author} {\bibinfo {author} {\bibfnamefont {D.}~\bibnamefont
  {Zhao}}\ and\ \bibinfo {author} {\bibfnamefont {H.}~\bibnamefont {Aluie}},\
  }\href@noop {} {\bibfield  {journal} {\bibinfo  {journal} {Physical Review
  Fluids}\ }\textbf {\bibinfo {volume} {3}},\ \bibinfo {pages} {301} (\bibinfo
  {year} {2018})}\BibitemShut {NoStop}%
\bibitem [{\citenamefont {Kraichnan}(1966)}]{Kraichnan66}%
  \BibitemOpen
  \bibfield  {author} {\bibinfo {author} {\bibfnamefont {R.~H.}\ \bibnamefont
  {Kraichnan}},\ }\href@noop {} {\bibfield  {journal} {\bibinfo  {journal}
  {Physics of Fluids}\ }\textbf {\bibinfo {volume} {9}},\ \bibinfo {pages}
  {1728} (\bibinfo {year} {1966})}\BibitemShut {NoStop}%
\bibitem [{\citenamefont {Matthaeus}\ and\ \citenamefont
  {Goldstein}(1982)}]{MatthaeusGoldstein82}%
  \BibitemOpen
  \bibfield  {author} {\bibinfo {author} {\bibfnamefont {W.~H.}\ \bibnamefont
  {Matthaeus}}\ and\ \bibinfo {author} {\bibfnamefont {M.~L.}\ \bibnamefont
  {Goldstein}},\ }\href@noop {} {\bibfield  {journal} {\bibinfo  {journal}
  {Journal of Geophysical Research}\ }\textbf {\bibinfo {volume} {87}},\
  \bibinfo {pages} {10347} (\bibinfo {year} {1982})}\BibitemShut {NoStop}%
\bibitem [{\citenamefont {Matthaeus}\ and\ \citenamefont
  {Lamkin}(1986)}]{MatthaeusLamkin86}%
  \BibitemOpen
  \bibfield  {author} {\bibinfo {author} {\bibfnamefont {W.~H.}\ \bibnamefont
  {Matthaeus}}\ and\ \bibinfo {author} {\bibfnamefont {S.~L.}\ \bibnamefont
  {Lamkin}},\ }\href@noop {} {\bibfield  {journal} {\bibinfo  {journal}
  {Physics of Fluids}\ }\textbf {\bibinfo {volume} {29}},\ \bibinfo {pages}
  {2513} (\bibinfo {year} {1986})}\BibitemShut {NoStop}%
\bibitem [{\citenamefont {Grappin}\ \emph {et~al.}(1983)\citenamefont
  {Grappin}, \citenamefont {Leorat},\ and\ \citenamefont
  {Pouquet}}]{Grappinetal83}%
  \BibitemOpen
  \bibfield  {author} {\bibinfo {author} {\bibfnamefont {R.}~\bibnamefont
  {Grappin}}, \bibinfo {author} {\bibfnamefont {J.}~\bibnamefont {Leorat}}, \
  and\ \bibinfo {author} {\bibfnamefont {A.}~\bibnamefont {Pouquet}},\
  }\href@noop {} {\bibfield  {journal} {\bibinfo  {journal} {Astronomy and
  Astrophysics}\ }\textbf {\bibinfo {volume} {126}},\ \bibinfo {pages} {51}
  (\bibinfo {year} {1983})}\BibitemShut {NoStop}%
\bibitem [{\citenamefont {Boldyrev}\ \emph {et~al.}(2011)\citenamefont
  {Boldyrev}, \citenamefont {Perez}, \citenamefont {Borovsky},\ and\
  \citenamefont {Podesta}}]{Boldyrevetal11}%
  \BibitemOpen
  \bibfield  {author} {\bibinfo {author} {\bibfnamefont {S.}~\bibnamefont
  {Boldyrev}}, \bibinfo {author} {\bibfnamefont {J.~C.}\ \bibnamefont {Perez}},
  \bibinfo {author} {\bibfnamefont {J.~E.}\ \bibnamefont {Borovsky}}, \ and\
  \bibinfo {author} {\bibfnamefont {J.~J.}\ \bibnamefont {Podesta}},\
  }\href@noop {} {\bibfield  {journal} {\bibinfo  {journal} {The Astrophysical
  Journal Letters}\ }\textbf {\bibinfo {volume} {741}},\ \bibinfo {pages} {L19}
  (\bibinfo {year} {2011})}\BibitemShut {NoStop}%
\bibitem [{\citenamefont {Zel'Dovich}\ \emph {et~al.}(1984)\citenamefont
  {Zel'Dovich}, \citenamefont {Ruzmaikin}, \citenamefont {Molchanov},\ and\
  \citenamefont {Sokoloff}}]{Zel'Dovichetal84}%
  \BibitemOpen
  \bibfield  {author} {\bibinfo {author} {\bibfnamefont {Y.~B.}\ \bibnamefont
  {Zel'Dovich}}, \bibinfo {author} {\bibfnamefont {A.}~\bibnamefont
  {Ruzmaikin}}, \bibinfo {author} {\bibfnamefont {S.}~\bibnamefont
  {Molchanov}}, \ and\ \bibinfo {author} {\bibfnamefont {D.}~\bibnamefont
  {Sokoloff}},\ }\href@noop {} {\bibfield  {journal} {\bibinfo  {journal}
  {Journal of Fluid Mechanics}\ }\textbf {\bibinfo {volume} {144}},\ \bibinfo
  {pages} {1} (\bibinfo {year} {1984})}\BibitemShut {NoStop}%
\bibitem [{\citenamefont {{Schekochihin}}\ \emph {et~al.}(2002)\citenamefont
  {{Schekochihin}}, \citenamefont {{Boldyrev}},\ and\ \citenamefont
  {{Kulsrud}}}]{Schekochihinetal02}%
  \BibitemOpen
  \bibfield  {author} {\bibinfo {author} {\bibfnamefont {A.~A.}\ \bibnamefont
  {{Schekochihin}}}, \bibinfo {author} {\bibfnamefont {S.~A.}\ \bibnamefont
  {{Boldyrev}}}, \ and\ \bibinfo {author} {\bibfnamefont {R.~M.}\ \bibnamefont
  {{Kulsrud}}},\ }\href {\doibase 10.1086/338697} {\bibfield  {journal}
  {\bibinfo  {journal} {Astrophys. J.}\ }\textbf {\bibinfo {volume} {567}},\
  \bibinfo {pages} {828} (\bibinfo {year} {2002})}\BibitemShut {NoStop}%
\bibitem [{\citenamefont {Aluie}(2011)}]{Aluie11}%
  \BibitemOpen
  \bibfield  {author} {\bibinfo {author} {\bibfnamefont {H.}~\bibnamefont
  {Aluie}},\ }\href@noop {} {\bibfield  {journal} {\bibinfo  {journal}
  {Physical Review Letters}\ }\textbf {\bibinfo {volume} {106}},\ \bibinfo
  {pages} {174502} (\bibinfo {year} {2011})}\BibitemShut {NoStop}%
\bibitem [{\citenamefont {Aluie}\ \emph {et~al.}(2012)\citenamefont {Aluie},
  \citenamefont {Li},\ and\ \citenamefont {Li}}]{Aluieetal12}%
  \BibitemOpen
  \bibfield  {author} {\bibinfo {author} {\bibfnamefont {H.}~\bibnamefont
  {Aluie}}, \bibinfo {author} {\bibfnamefont {S.}~\bibnamefont {Li}}, \ and\
  \bibinfo {author} {\bibfnamefont {H.}~\bibnamefont {Li}},\ }\href@noop {}
  {\bibfield  {journal} {\bibinfo  {journal} {The Astrophysical Journal}\
  }\textbf {\bibinfo {volume} {751}},\ \bibinfo {pages} {L29} (\bibinfo {year}
  {2012})}\BibitemShut {NoStop}%
\bibitem [{\citenamefont {Yang}\ \emph {et~al.}(2016)\citenamefont {Yang},
  \citenamefont {Shi}, \citenamefont {Wan}, \citenamefont {Matthaeus},\ and\
  \citenamefont {Chen}}]{Yangetal16}%
  \BibitemOpen
  \bibfield  {author} {\bibinfo {author} {\bibfnamefont {Y.}~\bibnamefont
  {Yang}}, \bibinfo {author} {\bibfnamefont {Y.}~\bibnamefont {Shi}}, \bibinfo
  {author} {\bibfnamefont {M.}~\bibnamefont {Wan}}, \bibinfo {author}
  {\bibfnamefont {W.~H.}\ \bibnamefont {Matthaeus}}, \ and\ \bibinfo {author}
  {\bibfnamefont {S.}~\bibnamefont {Chen}},\ }\href@noop {} {\bibfield
  {journal} {\bibinfo  {journal} {Physical Review E}\ }\textbf {\bibinfo
  {volume} {93}},\ \bibinfo {pages} {061102} (\bibinfo {year}
  {2016})}\BibitemShut {NoStop}%
\bibitem [{\citenamefont {Mason}\ \emph {et~al.}(2008)\citenamefont {Mason},
  \citenamefont {Cattaneo},\ and\ \citenamefont {Boldyrev}}]{Masonetal08}%
  \BibitemOpen
  \bibfield  {author} {\bibinfo {author} {\bibfnamefont {J.}~\bibnamefont
  {Mason}}, \bibinfo {author} {\bibfnamefont {F.}~\bibnamefont {Cattaneo}}, \
  and\ \bibinfo {author} {\bibfnamefont {S.}~\bibnamefont {Boldyrev}},\
  }\href@noop {} {\bibfield  {journal} {\bibinfo  {journal} {Physical review.
  E, Statistical, nonlinear, and soft matter physics}\ }\textbf {\bibinfo
  {volume} {77}},\ \bibinfo {pages} {742} (\bibinfo {year} {2008})}\BibitemShut
  {NoStop}%
\bibitem [{\citenamefont {Matthaeus}\ \emph {et~al.}(2008)\citenamefont
  {Matthaeus}, \citenamefont {Pouquet}, \citenamefont {Mininni}, \citenamefont
  {Dmitruk},\ and\ \citenamefont {Breech}}]{Matthaeusetal08}%
  \BibitemOpen
  \bibfield  {author} {\bibinfo {author} {\bibfnamefont {W.}~\bibnamefont
  {Matthaeus}}, \bibinfo {author} {\bibfnamefont {A.}~\bibnamefont {Pouquet}},
  \bibinfo {author} {\bibfnamefont {P.}~\bibnamefont {Mininni}}, \bibinfo
  {author} {\bibfnamefont {P.}~\bibnamefont {Dmitruk}}, \ and\ \bibinfo
  {author} {\bibfnamefont {B.}~\bibnamefont {Breech}},\ }\href@noop {}
  {\bibfield  {journal} {\bibinfo  {journal} {Physical review letters}\
  }\textbf {\bibinfo {volume} {100}},\ \bibinfo {pages} {085003} (\bibinfo
  {year} {2008})}\BibitemShut {NoStop}%
\bibitem [{\citenamefont {Krstulovic}\ \emph {et~al.}(2014)\citenamefont
  {Krstulovic}, \citenamefont {Brachet},\ and\ \citenamefont
  {Pouquet}}]{Krstulovicetal14}%
  \BibitemOpen
  \bibfield  {author} {\bibinfo {author} {\bibfnamefont {G.}~\bibnamefont
  {Krstulovic}}, \bibinfo {author} {\bibfnamefont {M.~E.}\ \bibnamefont
  {Brachet}}, \ and\ \bibinfo {author} {\bibfnamefont {A.}~\bibnamefont
  {Pouquet}},\ }\href@noop {} {\bibfield  {journal} {\bibinfo  {journal}
  {Physical review. E, Statistical, nonlinear, and soft matter physics}\
  }\textbf {\bibinfo {volume} {89}},\ \bibinfo {pages} {51} (\bibinfo {year}
  {2014})}\BibitemShut {NoStop}%
\bibitem [{\citenamefont {Politano}\ and\ \citenamefont
  {Pouquet}(1998{\natexlab{a}})}]{PolitanoPouquet98a}%
  \BibitemOpen
  \bibfield  {author} {\bibinfo {author} {\bibfnamefont {H.}~\bibnamefont
  {Politano}}\ and\ \bibinfo {author} {\bibfnamefont {A.}~\bibnamefont
  {Pouquet}},\ }\href@noop {} {\bibfield  {journal} {\bibinfo  {journal}
  {Physical Review E}\ }\textbf {\bibinfo {volume} {57}},\ \bibinfo {pages}
  {R21} (\bibinfo {year} {1998}{\natexlab{a}})}\BibitemShut {NoStop}%
\bibitem [{\citenamefont {Politano}\ and\ \citenamefont
  {Pouquet}(1998{\natexlab{b}})}]{PolitanoPouquet98b}%
  \BibitemOpen
  \bibfield  {author} {\bibinfo {author} {\bibfnamefont {H.}~\bibnamefont
  {Politano}}\ and\ \bibinfo {author} {\bibfnamefont {A.}~\bibnamefont
  {Pouquet}},\ }\href@noop {} {\bibfield  {journal} {\bibinfo  {journal}
  {Geophysical Research Letters}\ }\textbf {\bibinfo {volume} {25}},\ \bibinfo
  {pages} {273} (\bibinfo {year} {1998}{\natexlab{b}})}\BibitemShut {NoStop}%
\bibitem [{\citenamefont {{M{\"u}ller}}\ and\ \citenamefont
  {{Carati}}(2002)}]{MuellerCarati02}%
  \BibitemOpen
  \bibfield  {author} {\bibinfo {author} {\bibfnamefont {W.-C.}\ \bibnamefont
  {{M{\"u}ller}}}\ and\ \bibinfo {author} {\bibfnamefont {D.}~\bibnamefont
  {{Carati}}},\ }\href {\doibase 10.1063/1.1448498} {\bibfield  {journal}
  {\bibinfo  {journal} {Physics of Plasmas}\ }\textbf {\bibinfo {volume} {9}},\
  \bibinfo {pages} {824} (\bibinfo {year} {2002})}\BibitemShut {NoStop}%
\bibitem [{\citenamefont {Miesch}\ \emph {et~al.}(2015)\citenamefont {Miesch},
  \citenamefont {Matthaeus}, \citenamefont {Brandenburg}, \citenamefont
  {Petrosyan}, \citenamefont {Pouquet}, \citenamefont {Cambon}, \citenamefont
  {Jenko}, \citenamefont {Uzdensky}, \citenamefont {Stone}, \citenamefont
  {Tobias}, \citenamefont {Toomre},\ and\ \citenamefont
  {Velli}}]{Mieschetal15}%
  \BibitemOpen
  \bibfield  {author} {\bibinfo {author} {\bibfnamefont {M.}~\bibnamefont
  {Miesch}}, \bibinfo {author} {\bibfnamefont {W.}~\bibnamefont {Matthaeus}},
  \bibinfo {author} {\bibfnamefont {A.}~\bibnamefont {Brandenburg}}, \bibinfo
  {author} {\bibfnamefont {A.}~\bibnamefont {Petrosyan}}, \bibinfo {author}
  {\bibfnamefont {A.}~\bibnamefont {Pouquet}}, \bibinfo {author} {\bibfnamefont
  {C.}~\bibnamefont {Cambon}}, \bibinfo {author} {\bibfnamefont
  {F.}~\bibnamefont {Jenko}}, \bibinfo {author} {\bibfnamefont
  {D.}~\bibnamefont {Uzdensky}}, \bibinfo {author} {\bibfnamefont
  {J.}~\bibnamefont {Stone}}, \bibinfo {author} {\bibfnamefont
  {S.}~\bibnamefont {Tobias}}, \bibinfo {author} {\bibfnamefont
  {J.}~\bibnamefont {Toomre}}, \ and\ \bibinfo {author} {\bibfnamefont
  {M.}~\bibnamefont {Velli}},\ }\href@noop {} {\bibfield  {journal} {\bibinfo
  {journal} {Space Science Reviews}\ }\textbf {\bibinfo {volume} {194}},\
  \bibinfo {pages} {97} (\bibinfo {year} {2015})}\BibitemShut {NoStop}%
\bibitem [{\citenamefont {Lazarian}\ and\ \citenamefont
  {Vishniac}(1999)}]{LazarianVishniac99}%
  \BibitemOpen
  \bibfield  {author} {\bibinfo {author} {\bibfnamefont {A.}~\bibnamefont
  {Lazarian}}\ and\ \bibinfo {author} {\bibfnamefont {E.~T.}\ \bibnamefont
  {Vishniac}},\ }\href@noop {} {\bibfield  {journal} {\bibinfo  {journal}
  {Astrophysical Journal}\ }\textbf {\bibinfo {volume} {517}},\ \bibinfo
  {pages} {700} (\bibinfo {year} {1999})}\BibitemShut {NoStop}%
\bibitem [{\citenamefont {Eyink}\ \emph {et~al.}(2013)\citenamefont {Eyink},
  \citenamefont {Vishniac}, \citenamefont {Lalescu}, \citenamefont {Aluie},
  \citenamefont {Kanov}, \citenamefont {B{\"u}rger}, \citenamefont {Burns},
  \citenamefont {Meneveau},\ and\ \citenamefont {Szalay}}]{Eyinketal13}%
  \BibitemOpen
  \bibfield  {author} {\bibinfo {author} {\bibfnamefont {G.}~\bibnamefont
  {Eyink}}, \bibinfo {author} {\bibfnamefont {E.}~\bibnamefont {Vishniac}},
  \bibinfo {author} {\bibfnamefont {C.}~\bibnamefont {Lalescu}}, \bibinfo
  {author} {\bibfnamefont {H.}~\bibnamefont {Aluie}}, \bibinfo {author}
  {\bibfnamefont {K.}~\bibnamefont {Kanov}}, \bibinfo {author} {\bibfnamefont
  {K.}~\bibnamefont {B{\"u}rger}}, \bibinfo {author} {\bibfnamefont
  {R.}~\bibnamefont {Burns}}, \bibinfo {author} {\bibfnamefont
  {C.}~\bibnamefont {Meneveau}}, \ and\ \bibinfo {author} {\bibfnamefont
  {A.}~\bibnamefont {Szalay}},\ }\href@noop {} {\bibfield  {journal} {\bibinfo
  {journal} {Nature}\ }\textbf {\bibinfo {volume} {497}},\ \bibinfo {pages}
  {466} (\bibinfo {year} {2013})}\BibitemShut {NoStop}%
\bibitem [{\citenamefont {Eyink}(2015)}]{Eyink15}%
  \BibitemOpen
  \bibfield  {author} {\bibinfo {author} {\bibfnamefont {G.~L.}\ \bibnamefont
  {Eyink}},\ }\href@noop {} {\bibfield  {journal} {\bibinfo  {journal} {The
  Astrophysical Journal}\ }\textbf {\bibinfo {volume} {807}},\ \bibinfo {pages}
  {137} (\bibinfo {year} {2015})}\BibitemShut {NoStop}%
\bibitem [{\citenamefont {Eyink}\ and\ \citenamefont
  {Aluie}(2006)}]{EyinkAluie06}%
  \BibitemOpen
  \bibfield  {author} {\bibinfo {author} {\bibfnamefont {G.~L.}\ \bibnamefont
  {Eyink}}\ and\ \bibinfo {author} {\bibfnamefont {H.}~\bibnamefont {Aluie}},\
  }\href@noop {} {\bibfield  {journal} {\bibinfo  {journal} {Physica D:
  Nonlinear Phenomena}\ }\textbf {\bibinfo {volume} {223}},\ \bibinfo {pages}
  {82} (\bibinfo {year} {2006})}\BibitemShut {NoStop}%
\bibitem [{\citenamefont {Blackman}\ and\ \citenamefont
  {Field}(2002)}]{BlackmanField02}%
  \BibitemOpen
  \bibfield  {author} {\bibinfo {author} {\bibfnamefont {E.~G.}\ \bibnamefont
  {Blackman}}\ and\ \bibinfo {author} {\bibfnamefont {G.~B.}\ \bibnamefont
  {Field}},\ }\href@noop {} {\bibfield  {journal} {\bibinfo  {journal}
  {Physical Review Letters}\ }\textbf {\bibinfo {volume} {89}},\ \bibinfo
  {pages} {265007} (\bibinfo {year} {2002})}\BibitemShut {NoStop}%
\bibitem [{\citenamefont {Beresnyak}(2012)}]{Beresnyak12}%
  \BibitemOpen
  \bibfield  {author} {\bibinfo {author} {\bibfnamefont {A.}~\bibnamefont
  {Beresnyak}},\ }\href@noop {} {\bibfield  {journal} {\bibinfo  {journal}
  {Physical Review Letters}\ }\textbf {\bibinfo {volume} {108}},\ \bibinfo
  {pages} {1031} (\bibinfo {year} {2012})}\BibitemShut {NoStop}%
\bibitem [{\citenamefont {Brandenburg}(2018)}]{Brandenburg18}%
  \BibitemOpen
  \bibfield  {author} {\bibinfo {author} {\bibfnamefont {A.}~\bibnamefont
  {Brandenburg}},\ }\href {\doibase 10.1017/S0022377818000806} {\bibfield
  {journal} {\bibinfo  {journal} {Journal of Plasma Physics}\ }\textbf
  {\bibinfo {volume} {84}},\ \bibinfo {pages} {735840404} (\bibinfo {year}
  {2018})}\BibitemShut {NoStop}%
\bibitem [{\citenamefont {Offermans}\ \emph {et~al.}(2018)\citenamefont
  {Offermans}, \citenamefont {Biferale}, \citenamefont {Buzzicotti},\ and\
  \citenamefont {Linkmann}}]{Offermansetal18}%
  \BibitemOpen
  \bibfield  {author} {\bibinfo {author} {\bibfnamefont {G.~P.}\ \bibnamefont
  {Offermans}}, \bibinfo {author} {\bibfnamefont {L.}~\bibnamefont {Biferale}},
  \bibinfo {author} {\bibfnamefont {M.}~\bibnamefont {Buzzicotti}}, \ and\
  \bibinfo {author} {\bibfnamefont {M.}~\bibnamefont {Linkmann}},\ }\href@noop
  {} {\bibfield  {journal} {\bibinfo  {journal} {arXiv.org}\ ,\ \bibinfo
  {pages} {arXiv:1807.00759}} (\bibinfo {year} {2018})},\ \Eprint
  {http://arxiv.org/abs/1807.00759} {1807.00759} \BibitemShut {NoStop}%
\bibitem [{\citenamefont {Patterson}\ and\ \citenamefont
  {Orszag}(1971)}]{PattersonOrszag71}%
  \BibitemOpen
  \bibfield  {author} {\bibinfo {author} {\bibfnamefont {G.~S.}\ \bibnamefont
  {Patterson}}\ and\ \bibinfo {author} {\bibfnamefont {S.~A.}\ \bibnamefont
  {Orszag}},\ }\href@noop {} {\bibfield  {journal} {\bibinfo  {journal}
  {Physics of Fluids}\ }\textbf {\bibinfo {volume} {14}},\ \bibinfo {pages}
  {2538} (\bibinfo {year} {1971})}\BibitemShut {NoStop}%
\bibitem [{\citenamefont {Borue}\ and\ \citenamefont
  {Orszag}(1995)}]{BorueOrszag95}%
  \BibitemOpen
  \bibfield  {author} {\bibinfo {author} {\bibfnamefont {V.}~\bibnamefont
  {Borue}}\ and\ \bibinfo {author} {\bibfnamefont {S.~A.}\ \bibnamefont
  {Orszag}},\ }\href {http://stacks.iop.org/0295-5075/29/i=9/a=006} {\bibfield
  {journal} {\bibinfo  {journal} {EPL (Europhysics Letters)}\ }\textbf
  {\bibinfo {volume} {29}},\ \bibinfo {pages} {687} (\bibinfo {year}
  {1995})}\BibitemShut {NoStop}%
\bibitem [{\citenamefont {Cho}\ and\ \citenamefont
  {Vishniac}(2000)}]{ChoVishniac00}%
  \BibitemOpen
  \bibfield  {author} {\bibinfo {author} {\bibfnamefont {J.}~\bibnamefont
  {Cho}}\ and\ \bibinfo {author} {\bibfnamefont {E.~T.}\ \bibnamefont
  {Vishniac}},\ }\href {http://stacks.iop.org/0004-637X/539/i=1/a=273}
  {\bibfield  {journal} {\bibinfo  {journal} {The Astrophysical Journal}\
  }\textbf {\bibinfo {volume} {539}},\ \bibinfo {pages} {273} (\bibinfo {year}
  {2000})}\BibitemShut {NoStop}%
\bibitem [{\citenamefont {Kawai}(2013)}]{Kawai13}%
  \BibitemOpen
  \bibfield  {author} {\bibinfo {author} {\bibfnamefont {S.}~\bibnamefont
  {Kawai}},\ }\href {\doibase 10.1016/j.jcp.2013.05.033} {\bibfield  {journal}
  {\bibinfo  {journal} {Journal of Computational Physics}\ }\textbf {\bibinfo
  {volume} {251}},\ \bibinfo {pages} {292} (\bibinfo {year}
  {2013})}\BibitemShut {NoStop}%
\bibitem [{\citenamefont {Beresnyak}(2015)}]{Beresnyak15}%
  \BibitemOpen
  \bibfield  {author} {\bibinfo {author} {\bibfnamefont {A.}~\bibnamefont
  {Beresnyak}},\ }\href@noop {} {\bibfield  {journal} {\bibinfo  {journal} {The
  Astrophysical Journal Letters}\ }\textbf {\bibinfo {volume} {801}},\ \bibinfo
  {pages} {L9} (\bibinfo {year} {2015})}\BibitemShut {NoStop}%
\bibitem [{\citenamefont {Meyrand}\ \emph {et~al.}(2016)\citenamefont
  {Meyrand}, \citenamefont {Galtier},\ and\ \citenamefont
  {Kiyani}}]{Meyrandetal16}%
  \BibitemOpen
  \bibfield  {author} {\bibinfo {author} {\bibfnamefont {R.}~\bibnamefont
  {Meyrand}}, \bibinfo {author} {\bibfnamefont {S.}~\bibnamefont {Galtier}}, \
  and\ \bibinfo {author} {\bibfnamefont {K.~H.}\ \bibnamefont {Kiyani}},\
  }\href@noop {} {\bibfield  {journal} {\bibinfo  {journal} {Physical review
  letters}\ }\textbf {\bibinfo {volume} {116}},\ \bibinfo {pages} {105002}
  (\bibinfo {year} {2016})}\BibitemShut {NoStop}%
\bibitem [{\citenamefont {Kawazura}\ \emph {et~al.}(2018)\citenamefont
  {Kawazura}, \citenamefont {Barnes},\ and\ \citenamefont
  {Schekochihin}}]{Kawazuraetal18}%
  \BibitemOpen
  \bibfield  {author} {\bibinfo {author} {\bibfnamefont {Y.}~\bibnamefont
  {Kawazura}}, \bibinfo {author} {\bibfnamefont {M.}~\bibnamefont {Barnes}}, \
  and\ \bibinfo {author} {\bibfnamefont {A.~A.}\ \bibnamefont {Schekochihin}},\
  }\href@noop {} {\bibfield  {journal} {\bibinfo  {journal} {arXiv}\ ,\
  \bibinfo {pages} {arXiv:1807.07702}} (\bibinfo {year} {2018})}\BibitemShut
  {NoStop}%
\bibitem [{\citenamefont {Biskamp}\ and\ \citenamefont
  {M{\"u}ller}(2000)}]{BiskampMuller00}%
  \BibitemOpen
  \bibfield  {author} {\bibinfo {author} {\bibfnamefont {D.}~\bibnamefont
  {Biskamp}}\ and\ \bibinfo {author} {\bibfnamefont {W.-C.}\ \bibnamefont
  {M{\"u}ller}},\ }\href@noop {} {\bibfield  {journal} {\bibinfo  {journal}
  {Physics of Plasmas}\ }\textbf {\bibinfo {volume} {7}},\ \bibinfo {pages}
  {4889} (\bibinfo {year} {2000})}\BibitemShut {NoStop}%
\bibitem [{\citenamefont {Frisch}\ \emph {et~al.}(2008)\citenamefont {Frisch},
  \citenamefont {KURIEN}, \citenamefont {Pandit}, \citenamefont {Pauls},
  \citenamefont {Ray}, \citenamefont {Wirth},\ and\ \citenamefont
  {Zhu}}]{Frischetal08}%
  \BibitemOpen
  \bibfield  {author} {\bibinfo {author} {\bibfnamefont {U.}~\bibnamefont
  {Frisch}}, \bibinfo {author} {\bibfnamefont {S.}~\bibnamefont {KURIEN}},
  \bibinfo {author} {\bibfnamefont {R.}~\bibnamefont {Pandit}}, \bibinfo
  {author} {\bibfnamefont {W.}~\bibnamefont {Pauls}}, \bibinfo {author}
  {\bibfnamefont {S.~S.}\ \bibnamefont {Ray}}, \bibinfo {author} {\bibfnamefont
  {A.}~\bibnamefont {Wirth}}, \ and\ \bibinfo {author} {\bibfnamefont {J.-Z.}\
  \bibnamefont {Zhu}},\ }\href@noop {} {\bibfield  {journal} {\bibinfo
  {journal} {Physical Review Letters}\ }\textbf {\bibinfo {volume} {101}}
  (\bibinfo {year} {2008})}\BibitemShut {NoStop}%
\bibitem [{\citenamefont {Spyksma}\ \emph {et~al.}(2012)\citenamefont
  {Spyksma}, \citenamefont {Magcalas},\ and\ \citenamefont
  {Campbell}}]{Spyksmaetal12}%
  \BibitemOpen
  \bibfield  {author} {\bibinfo {author} {\bibfnamefont {K.}~\bibnamefont
  {Spyksma}}, \bibinfo {author} {\bibfnamefont {M.}~\bibnamefont {Magcalas}}, \
  and\ \bibinfo {author} {\bibfnamefont {N.}~\bibnamefont {Campbell}},\
  }\href@noop {} {\bibfield  {journal} {\bibinfo  {journal} {Physics of
  Fluids}\ }\textbf {\bibinfo {volume} {24}},\ \bibinfo {pages} {125102}
  (\bibinfo {year} {2012})}\BibitemShut {NoStop}%
\bibitem [{\citenamefont {Li}\ \emph {et~al.}(2016)\citenamefont {Li},
  \citenamefont {Howes}, \citenamefont {Klein},\ and\ \citenamefont
  {TenBarge}}]{Lietal16}%
  \BibitemOpen
  \bibfield  {author} {\bibinfo {author} {\bibfnamefont {T.~C.}\ \bibnamefont
  {Li}}, \bibinfo {author} {\bibfnamefont {G.~G.}\ \bibnamefont {Howes}},
  \bibinfo {author} {\bibfnamefont {K.~G.}\ \bibnamefont {Klein}}, \ and\
  \bibinfo {author} {\bibfnamefont {J.~M.}\ \bibnamefont {TenBarge}},\
  }\href@noop {} {\bibfield  {journal} {\bibinfo  {journal} {The Astrophysical
  Journal Letters}\ }\textbf {\bibinfo {volume} {832}},\ \bibinfo {pages} {L24}
  (\bibinfo {year} {2016})}\BibitemShut {NoStop}%
\bibitem [{\citenamefont {Braginskii}(1965)}]{Braginskii65}%
  \BibitemOpen
  \bibfield  {author} {\bibinfo {author} {\bibfnamefont {S.}~\bibnamefont
  {Braginskii}},\ }\href@noop {} {\emph {\bibinfo {title} {Transport processes
  in a plasma}}},\ Vol.~\bibinfo {volume} {1}\ (\bibinfo  {publisher}
  {Consultants Bureau, New York},\ \bibinfo {year} {1965})\BibitemShut
  {NoStop}%
\bibitem [{\citenamefont {Haines}(1986)}]{Haines86}%
  \BibitemOpen
  \bibfield  {author} {\bibinfo {author} {\bibfnamefont {M.~G.}\ \bibnamefont
  {Haines}},\ }\href@noop {} {\bibfield  {journal} {\bibinfo  {journal} {Plasma
  Physics and Controlled Fusion}\ }\textbf {\bibinfo {volume} {28}},\ \bibinfo
  {pages} {1705} (\bibinfo {year} {1986})}\BibitemShut {NoStop}%
\bibitem [{\citenamefont {Davies}\ \emph {et~al.}(2015)\citenamefont {Davies},
  \citenamefont {Betti}, \citenamefont {Chang},\ and\ \citenamefont
  {Fiksel}}]{Daviesetal15}%
  \BibitemOpen
  \bibfield  {author} {\bibinfo {author} {\bibfnamefont {J.~R.}\ \bibnamefont
  {Davies}}, \bibinfo {author} {\bibfnamefont {R.}~\bibnamefont {Betti}},
  \bibinfo {author} {\bibfnamefont {P.~Y.}\ \bibnamefont {Chang}}, \ and\
  \bibinfo {author} {\bibfnamefont {G.}~\bibnamefont {Fiksel}},\ }\href@noop {}
  {\bibfield  {journal} {\bibinfo  {journal} {Physics of Plasmas}\ }\textbf
  {\bibinfo {volume} {22}},\ \bibinfo {pages} {112703} (\bibinfo {year}
  {2015})}\BibitemShut {NoStop}%
\bibitem [{\citenamefont {Eyink}\ and\ \citenamefont
  {Aluie}(2009)}]{EyinkAluie09}%
  \BibitemOpen
  \bibfield  {author} {\bibinfo {author} {\bibfnamefont {G.~L.}\ \bibnamefont
  {Eyink}}\ and\ \bibinfo {author} {\bibfnamefont {H.}~\bibnamefont {Aluie}},\
  }\href@noop {} {\bibfield  {journal} {\bibinfo  {journal} {Physics of
  Fluids}\ }\textbf {\bibinfo {volume} {21}},\ \bibinfo {pages} {115107}
  (\bibinfo {year} {2009})}\BibitemShut {NoStop}%
\bibitem [{\citenamefont {Aluie}\ and\ \citenamefont
  {Eyink}(2009)}]{AluieEyink09}%
  \BibitemOpen
  \bibfield  {author} {\bibinfo {author} {\bibfnamefont {H.}~\bibnamefont
  {Aluie}}\ and\ \bibinfo {author} {\bibfnamefont {G.~L.}\ \bibnamefont
  {Eyink}},\ }\href@noop {} {\bibfield  {journal} {\bibinfo  {journal} {Physics
  of Fluids}\ }\textbf {\bibinfo {volume} {21}},\ \bibinfo {pages} {115108}
  (\bibinfo {year} {2009})}\BibitemShut {NoStop}%
\bibitem [{\citenamefont {Sadek}\ and\ \citenamefont
  {Aluie}(2018)}]{SadekAluie18}%
  \BibitemOpen
  \bibfield  {author} {\bibinfo {author} {\bibfnamefont {M.}~\bibnamefont
  {Sadek}}\ and\ \bibinfo {author} {\bibfnamefont {H.}~\bibnamefont {Aluie}},\
  }\href@noop {} {\bibfield  {journal} {\bibinfo  {journal} {Phys. Rev.
  Fluids}\ }\textbf {\bibinfo {volume} {3}},\ \bibinfo {pages} {124610}
  (\bibinfo {year} {2018})}\BibitemShut {NoStop}%
\bibitem [{\citenamefont {Frisch}(1995)}]{Frisch95}%
  \BibitemOpen
  \bibfield  {author} {\bibinfo {author} {\bibfnamefont {U.}~\bibnamefont
  {Frisch}},\ }\href@noop {} {\emph {\bibinfo {title} {{{T}urbulence: the
  legacy of {A}. {N}. {K}olmogorov}}}}\ (\bibinfo  {publisher} {Cambridge
  University Press, UK},\ \bibinfo {year} {1995})\BibitemShut {NoStop}%
\end{thebibliography}%

\end{document}